%% file: 0paper.tex
\documentclass[%
 reprint,
 amsmath,amssymb,
 aps,
 showkeys,
]{revtex4-2}

\usepackage{dcolumn}
\usepackage{lipsum}  
\usepackage{graphicx}
\usepackage{floatflt,epsfig} 
\usepackage{mathtools}
\usepackage{color}
\usepackage{braket}
\usepackage{hyperref}
\usepackage[normalem]{ulem}
\usepackage[english]{babel}
\usepackage{blindtext}
\usepackage{lipsum}  
\usepackage{amsthm}
\usepackage{tabularx}
\usepackage{bm}
\usepackage{dsfont}
\usepackage{bbold}
\usepackage[percent]{overpic}
\usepackage{cmupint}
\usepackage{caption, subcaption}
\usepackage{sidecap}

\usepackage[newcommands]{ragged2e} 
\usepackage{subcaption}
\newcommand {\br} {\mathbf{r}}

\newcommand {\bk} {\mathbf{k}}
\newcommand {\bJ} {\mathbf{J}}
\newcommand {\bA} {\mathbf{A}}

\newcommand {\bK} {\mathbf{K}}

\newcommand {\bepsilon} {\bm{\epsilon}}
\newcommand {\dphan} {\vphantom{\dagger}}

\newcommand{\Jp}{\mathbf{J}_\mathrm{p}}
\newcommand{\hatJp}{\hat{\mathbf{J}}_\mathrm{p}}
\newcommand{\hatjp}{\hat{\mathbf{j}}_\mathrm{p}}

\newcommand{\hatJd}{\hat{\mathbf{J}}_\mathrm{d}}

\newcommand{\Fourier}{\mathcal{F}}
\newcommand{\omda}{\omega_{\mathrm{d},\alpha}}
\newcommand{\rmd}{\mathrm{d}}
\newcommand{\rme}{\mathrm{e}}
\newcommand{\kF}{k_\mathrm{F}}
\newcommand{\Conv}{
  \mathop{\scalebox{1.5}{\raisebox{-0.2ex}{$\circledast$}}
  }
}
\newcommand {\obr} {\overline{\mathbf{r}}}

\newcommand{\FF}{\mathbf{F}}
\newcommand{\imagi}{\mathrm{i}}

\newcommand {\bj} {\mathbf{j}}

\begin{document}


\title{Making \textit{ab initio} QED functional(s): Non-perturbative and photon-free effective frameworks for strong light-matter coupling}

  \author{Christian Sch\"afer$^{1,2,3,4}$}
  \email[Electronic address:\;]{christian.schaefer.physics@gmail.com}
  \author{Florian Buchholz$^{1}$}
  \email[Electronic address:\;]{florian.buchholz@alumni.tu-berlin.de}
  \author{Markus Penz$^{5}$}
  \email[Electronic address:\;]{markus.penz@uibk.ac.at}
  \author{Michael Ruggenthaler$^{1,2}$}
  \email[Electronic address:\;]{michael.ruggenthaler@mpsd.mpg.de}
  \author{Angel Rubio$^{1,2}$}
  \email[Electronic address:\;]{angel.rubio@mpsd.mpg.de}
  \affiliation{$^1$ Max Planck Institute for the Structure and Dynamics of Matter and Center for Free-Electron Laser Science \& Department of Physics, Luruper Chaussee 149, 22761 Hamburg, Germany\\
  $^2$ The Hamburg Center for Ultrafast Imaging, Luruper Chaussee 149, 22761 Hamburg, Germany\\
  $^3$Department of Physics, Chalmers University of Technology, 412 96 G\"oteborg, Sweden\\
  $^4$Department of Microtechnology and Nanoscience, MC2, Chalmers University of Technology, 412 96 G\"oteborg, Sweden\\
  $^5$Department of Mathematics, University of Innsbruck, Technikerstra{\ss}e 13/7, A-6020 Innsbruck, Austria}

\date{\today}

\begin{abstract}
Strong light-matter coupling provides a promising path for the control of quantum matter where the latter is routinely described from first-principles. However, combining the quantized nature of light with this \textit{ab initio} tool set is challenging and merely developing, as the coupled light-matter Hilbert space is conceptually different and computational cost quickly becomes overwhelming.
In this work, we provide a non-perturbative photon-free formulation of quantum electrodynamics (QED) in the long-wavelength limit, which is formulated solely on the matter Hilbert space and can serve as an accurate starting point for such \textit{ab initio} methods. The present formulation is an extension of quantum mechanics that recovers the exact results of QED for the zero- and infinite-coupling limit, the infinite-frequency as well as the homogeneous limit and we can constructively increase its accuracy.
We show how this formulation can be used to devise approximations for quantum-electrodynamical density-functional theory (QEDFT), which in turn also allows to extend the ansatz to the full minimal-coupling problem and to non-adiabatic situations. Finally, we provide a simple local-density-type functional that takes the strong coupling to the transverse photon-degrees of freedom into account and includes the correct frequency and polarization dependence. This is the first QEDFT functional that accounts for the quantized nature of light while remaining computationally simple enough to allow its application to a large range of systems. All approximations allow the seamless application to periodic systems.
\end{abstract}

\maketitle

\section{Introduction}
In the last decade seminal experimental results~\cite{ebbesen2016,chikkaraddy2016,kockum2019ultrastrong,Basov2021} 
have demonstrated that the properties and dynamics of atoms, molecules and solids can be substantially modified by coupling strongly to the modes of a photonic environment. The strong coupling between light and matter in theses cases leads to the emergence of hybrid light-matter states (polaritons) which subsequently can be used to control, for instance, chemical reactions~\cite{hutchison2012,thomas2016,Munkhbat2018,galego2015,herrera2016cavity,groenhof2019tracking,schafer2021shining}, enhance charge and energy transport~\cite{coles2014b,orgiu2015,zhong2017energy,schachenmayer2015,feist2015,saez2018organic,du2018theory,schafer2019modification}, even over very large distances, and there are recent indications that they might be used to increase the critical temperature of superconductors \cite{thomas2019exploringsuperconductivity,sentef2018,schlawin2019cavity}. Many of these changes persist at normal ambient conditions and are hence promising for quantum-technological applications~\cite{ebbesen2016}. 
They question however our common perception that light and matter are distinct physical entities and call for a more unified description.

Typically, in such strong light-matter coupling situations only a few photonic modes, which are supported by a cavity geometry, play a substantial role. Their interaction with matter is however greatly enhanced in comparison to the free-space situation. Effective subwavelength confinement~\cite{chikkaraddy2016,baranov2017novel,forn2019ultrastrong,mueller2020deep,hertzog2021enhancing} or circuit geometries~\cite{goban2015,langford2017experimentally,turschmann2019coherent} enable hybridization strengths on the order of the matter excitation. A noticeable hybridization between light and matter (strong coupling) results in the emergence of polaritons in the excited states. The coupling between (artificial) cavity modes and matter excitation can be that considerable that all eigenvalues of the individual constituents are affected and even the ground state becomes correlated (ultra-strong coupling). For even larger hybridization, the character of light and matter becomes truly interlacing (deep ultra-strong coupling).

\begin{figure}
    \centering
    \includegraphics[width=0.99\columnwidth]{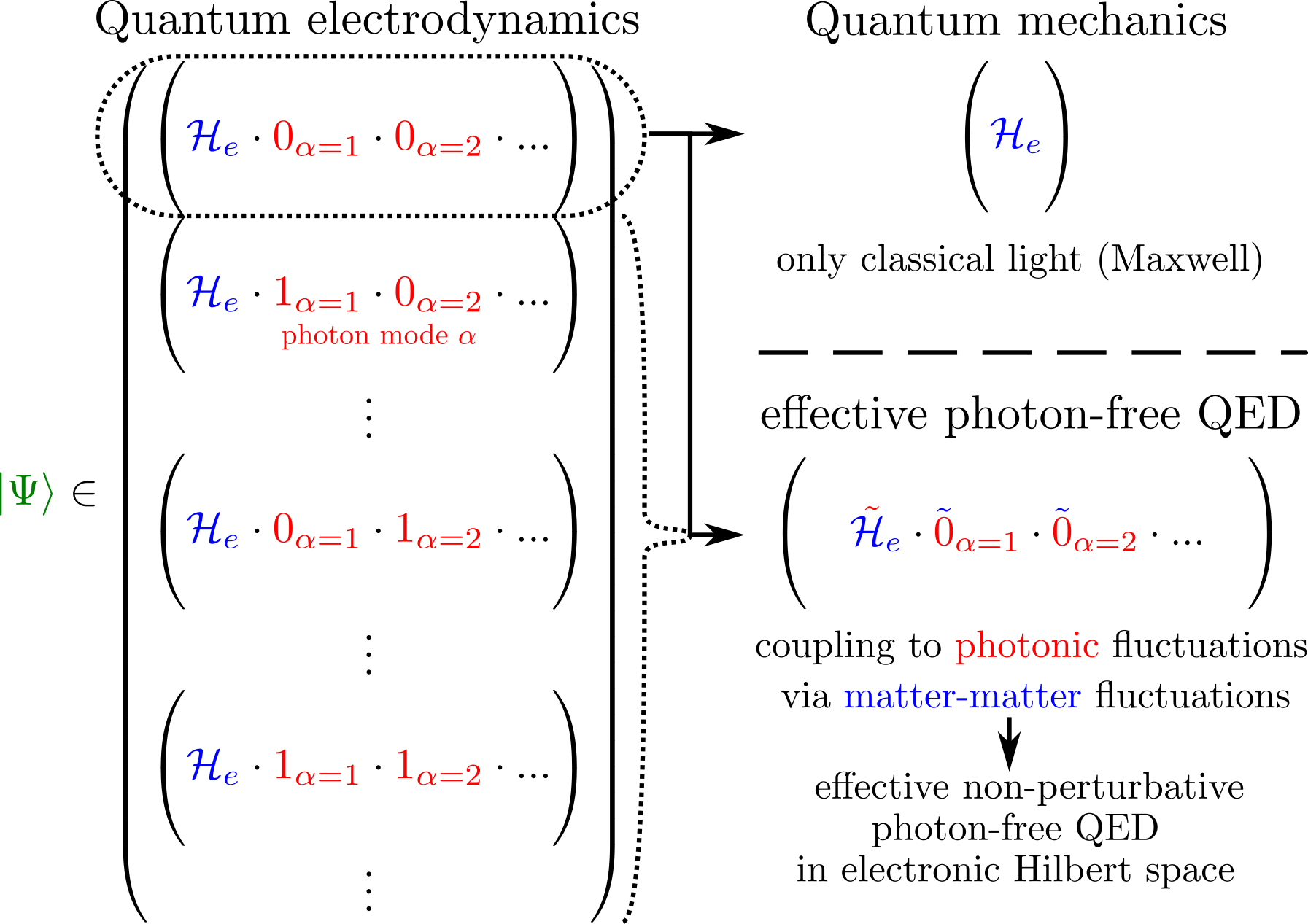}
    \caption{To the left, an illustration of the exponential increase in computational complexity: In QED the matter Hilbert space $\mathcal{H}_e$ is extended with the photonic Fock space (indicated by the vectors for different modes $\alpha$ starting with the vacuum state $0_{\alpha=1} \cdot 0_{\alpha=2} \cdot \dots$) and hence the combined Hilbert space grows exponentially also with the number of photonic states. A vast reduction of complexity is found by approximating QED with quantum mechanics, where the photonic sector is subsumed into the effective mass of the particles and into classical electromagnetic fields. The effective photon-free QED ansatz takes the dominant fluctuations of the quantized electromagnetic field $\tilde{0}$ into account, while having the same dimensionality as ordinary quantum mechanics.}
    \label{fig:concept}
\end{figure}

Many of the experimentally observed effects, in particular those involving chemical reactions under strong coupling, are so far theoretically not well understood~\cite{thomas2016,campos2020polaritonic,climent2020sn,li2021cavity,schafer2021shining}. Detailed theoretical explanations are largely missing as strong light-matter coupling calls in principle for investigations within the framework of quantum electrodynamics (QED)~\cite{cohen1997photons}. This, however, leads to an enormous increase of computational complexity as the combined light-matter Hilbert space becomes prohibitively large (illustrated in Fig.~\ref{fig:concept}).
The most common approach is to considerably reduce the dimensionality of the light-matter Hilbert space by deciding \textit{a priori} which matter and photon states are supposed to be important. This leads to effective model light-matter Hamiltonians~\cite{garraway2011,kockum2019ultrastrong}. A different approach is to reformulate the full QED problem in terms of reduced quantities. Quantum mechanics, for instance, can be viewed as a reduction of the full QED problem onto the matter sector only, where the photon field is taken into account approximately by the effective (physical) mass of the particles~\cite{cohen1997photons,craig1998} and by the longitudinal Coulomb interaction as well as by possible Van-der-Waals corrections~\cite{cohen1997photons,craig1998,haugland2020intermolecular}. This can be viewed as an example of an effective \textit{ab initio} light-matter Hamiltonian. This simplification is, however, no longer valid in the case of strong light-matter coupling and more advanced reformulations of QED become necessary. An in principle exact reformulation of QED on the basis of reduced quantities is quantum-electrodynamical density-functional theory (QEDFT), which allows to avoid the unfeasible coupled matter-photon wave function altogether~\cite{ruggenthaler2014,tokatly2013,ruggenthaler2015,jestadt2018real}. The main drawback of QEDFT is that we need to find approximations to the matter-photon coupling, usually in terms of matter quantities only~\cite{pellegrini2015,flick2017c}. Deriving such functional expressions for interactions is already for matter-only DFT a challenging task and the photonic field introduces additional and unfamiliar components into the problem. It therefore becomes desirable to define an effective \textit{ab initio} photon-free QED as an alternative to standard quantum mechanics which is also applicable to strong light-matter coupling situations and serves as a starting point for approximations for QEDFT. Furthermore, this effective matter-only theory should be constructively improvable and recover the physical limits of the original QED solution.

In this work we provide such a non-perturbative photon-free QED reformulation that takes the dominant fluctuations of the quantized electromagnetic field explicitly into account (Sec.~\ref{sec:ansatz}). It thereby lifts the artificial distinction between light and matter that underlies standard quantum mechanics and it remains applicable from the weak to the deep ultra-strong coupling regime.  Among other things we show that the ansatz recovers the exact QED solution in the weak- and infinite-coupling limit, the infinite-frequency limit, as well as for the homogeneous electron gas, without the need to treat the photonic degrees of freedom at all.
While many effective \textit{ab initio} descriptions of light-matter coupling problems are known, e.g., the high-frequency limit of Floquet theory~\cite{shirley1965solution,schafer2018insights,de2019floquet}, this high consistency with fundamental physical conditions provides a much more flexible and general perspective.
Since the theory is build upon an explicit ansatz for photonic operators in terms of matter quantities, we still have approximate access to photonic observables. Further we show that the non-perturbative photon-free QED reformulation is obtained as an truncation of a especially efficient basis-expansion of the full QED problem (Sec.~\ref{sec:pheg}). In this way, we can consistently increase the accuracy of the ansatz, converging to the original QED results and accounting for all correlations between light and matter. Furthermore, we show how this ansatz can be used to set up an orbital-dependent approximation in QEDFT that shares the same beneficial features as the non-perturbative photon-free QED reformulation (Sec.~\ref{sec:functionals}). Based on this functional we propose a simple local-density-type approximation for strong light-matter coupling that provides up to ultra-strong coupling excellent results and yet accounts for the correct frequency and cavity-polarization dependence. Due to the flexibility of the QEDFT framework, we can finally discuss how this ansatz can be extended to full minimal coupling and comment on non-adiabatic extensions. 

\section{Photon-free QED ansatz}\label{sec:ansatz}

The point of departure for our endeavour is the Pauli-Fierz (or minimal coupling) Hamiltonian of non-relativistic QED in Coulomb gauge. For simplicity, we will focus here at first on the long-wavelength limit, but the following conceptions are general enough to allow an abstraction beyond this common simplification. We provide an outlook beyond those common approximations in Sec.~\ref{sec:functionals}. In atomic units with electron charge $q = -\vert e \vert=-1$ and keeping the nuclei fixed we have
\begin{align}
\begin{split}
    \hat{H}_\mathrm{PF} &= \frac{1}{2} \sum_{i=1}^{N_e} \left( -\imagi\nabla_i + \frac{1}{c}\hat{\textbf{A}} \right)^2+ \sum_{i=1}^{N_e} v(\br_i) \\ &+ \frac{1}{2}\sum_{i\neq j}^{N_e} w(\br_i,\br_j) + \sum_{\alpha=1}^{M_p} \omega_\alpha \left(\hat{a}_\alpha^\dagger \hat{a}_\alpha + \frac{1}{2}\right)~.
    \label{eq:PF_Hamiltonian}
\end{split}
\end{align}
The $N_e$ electrons move according to the forces caused by the local nuclear potentials $v$ in addition to their 
longitudinal interaction $w$ and the coupling to the transverse vector potential at the molecular center-of-charge $\br_0$
\begin{align}
\label{eq:A_starting_def}
    \hat{\textbf{A}} = \sqrt{4\pi c^2} \sum_{\alpha=1}^{M_p} S_\alpha(\br_0) \bepsilon_\alpha \frac{1}{\sqrt{2\omega_\alpha}} \left(\hat{a}^\dagger_\alpha + \hat{a}_\alpha\right)~.
\end{align}
The $M_p$ cavity eigenmodes $S_\alpha(\br_0) \propto 1/\sqrt{V}$ are the solutions to the boundary value problem of the effective cavity (which we assume to be lossless for simplicity) and provide the local (vacuum) field-strength. The vector potential is conveniently expanded in the eigenmodes of the cavity geometry $\hat{\textbf{A}}=\sum_{\alpha} \hat{A}_\alpha \bepsilon_\alpha$, featuring polarization $\bepsilon_\alpha$, frequency $\omega_\alpha$ and cavity volume $V$. The $\hat{A}_\alpha$ obey the following equation of motion in the Heisenberg picture,
\begin{align}
    \left( \frac{1}{c^2}\frac{\rmd^2}{\rmd t^2} + \frac{\omega_\alpha^2}{c^2} \right) \hat{A}_{\alpha,{\rm H}} = -\frac{\lambda_\alpha^2}{c} \bepsilon_\alpha \cdot  \left(\hat{\mathbf{J}}_{\rm p,H} - \hat{\mathbf{J}}_{\rm d,H} \right),
    \label{eq:heisenberg}
\end{align}
with $\lambda_\alpha = \sqrt{4\pi}S_\alpha(\br_0)$ the fundamental coupling strength.
The paramagnetic current $\hatJp = -\imagi \sum_{i} \nabla_i$ serves as a driving force for the photonic degrees of freedom. The diamagnetic current in the long-wavelength approximation $\hatJd = -N_e/c \sum_{\alpha}\hat{A}_\alpha \bepsilon_\alpha$, on the other hand, can be conveniently absorbed in effective cavity frequencies and polarizations denoted from here on as $\tilde{\omega}_\alpha$ and $\tilde\bepsilon_\alpha$.
For a single mode, we can simply move $\hatJd$ to the left-hand-side to obtain $\tilde{\omega}_\alpha^2 = \omega_\alpha^2 + \omda^2$, $\omda^2 = N_e \lambda_\alpha^2$, and $\bepsilon_\alpha=\tilde\bepsilon_\alpha$. The diamagnetic term therefore dresses the bare cavity frequency by the fundamental coupling strength $\lambda_\alpha$ and the amount of charged particles. 
For multiple modes, a normal-mode (or Bogoliubov) transformation (see App.~\ref{App:bogo}) similarly eliminates the diamagnetic current, leading to a pure bilinear coupling in the Hamiltonian Eq.~\eqref{eq:PF_Hamiltonian}.
The representation in normal modes with $\tilde\omega_\alpha$ also transforms all related operators $\hat{A}$ where we keep the original notation for brevity.

After absorbing the diamagnetic current, the Ehrenfest, or more specifically Maxwell's, equation associated to Eq.~\eqref{eq:heisenberg} is easily solved with the help of the classical Greens function for the harmonic oscillator,
\begin{align}
    A_{\alpha} (t) = -\tilde{\omega}_\alpha\int_{-\infty}^t \rmd t' c \frac{\omda^2}{N_e\tilde{\omega}_\alpha^2} \sin\left(\tilde{\omega}_\alpha(t-t')\right) \tilde\bepsilon_\alpha \cdot \Jp(t')~.
    \label{eq:maxwell}
\end{align}
In the limit of Maxwell's equation there is therefore no need to keep track of the photonic degrees of freedom, as their evolution is fully determined by the initial conditions and the matter degrees of freedom.

\begin{figure*}
    \centering
    \includegraphics[width=1.0\textwidth]{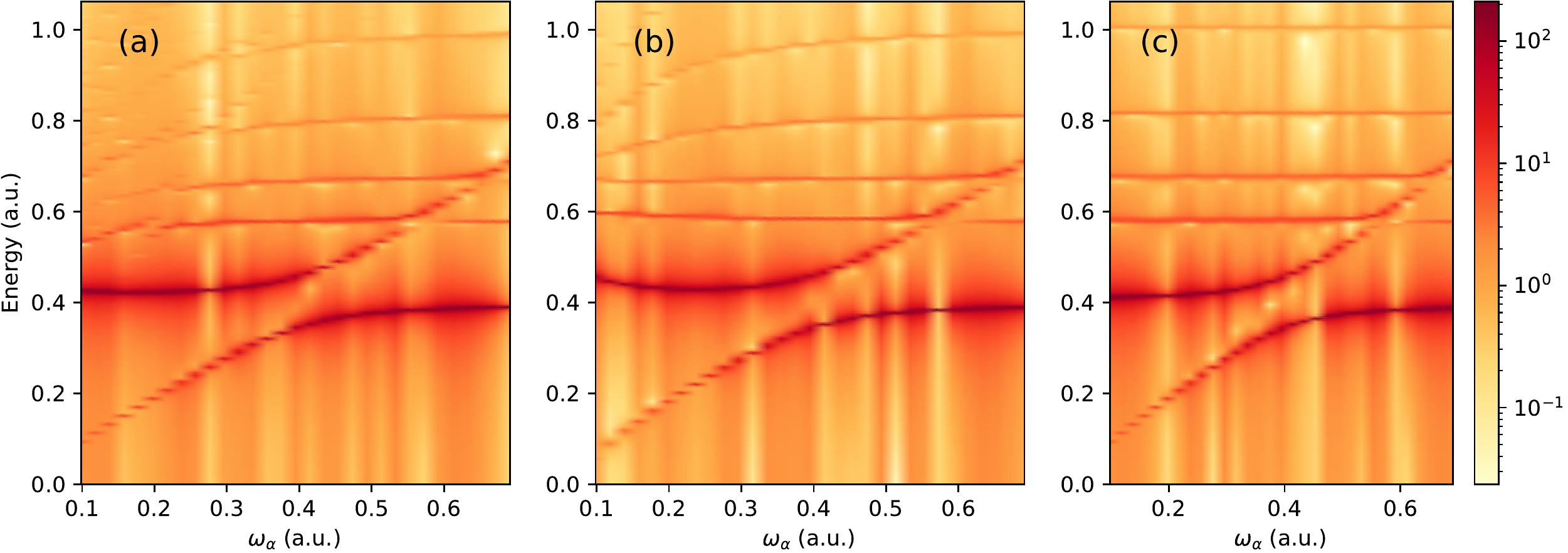}
    \caption{Linear-response spectrum of the dipole moment $\br(\omega_\alpha)$ for a single cavity-mode with different cavity frequencies $\omega_\alpha$ coupled to one-dimensional soft-Coulomb hydrogen $v(x) = -1/\sqrt{x^2+1}$. Shown are the exact reference (a, left), the photon-free effective Hamiltonian Eq.~\eqref{eq:breit} (b, middle), and the classical Maxwell solution (c, right). The fundamental coupling strength was chosen such that $g_\alpha/\omega_\alpha = 0.136,~ g_\alpha = \sqrt{\omega_\alpha/2}\lambda_\alpha$, i.e., on the interface between strong and ultra-strong coupling, with $\omega_\alpha = \varepsilon_{1}^{\text{hydrogen}}-\varepsilon_{0}^{\text{hydrogen}}$ in resonance with the first electronic excitation. We have chosen a grid of 301 points with $0.1~a_0$  spacing ($a_0$ being the Bohr radius) and 40 photonic Fock states for the exact reference using the PZW Hamiltonian Eq.~\eqref{eq:PZW_Hamiltonian}. The response was obtained by the delta-kick method applying a perturbation $v_{\text{kick}}(x,t)=-10^{-4}/\pi \cdot 10^{-2}/[(t-1)^2 + 10^{-4}] x$ 
    with consecutive time-propagation for $T=10^{3}~a.u.$ with 4th-order Runge-Kutta and a time-stepping of $dt=5\cdot10^{-4}~a.u.$}
    \label{fig:spectrumHB}
\end{figure*}

Now let us stretch this idea, comparably to the relativistic Breit substitution~\cite{PhysRev.39.616}, to the Heisenberg equation of motion and let us attempt to avoid the necessity of photonic degrees of freedom in Eq.~\eqref{eq:PF_Hamiltonian}. Certainly, such a substitution is far less trivial for the operator quantity, it is furthermore complicated by the fact that now we deal with two different pictures, i.e.,  Heisenberg for Eq.~\eqref{eq:heisenberg} and Schr\"odinger for Eq.~\eqref{eq:PF_Hamiltonian}. A suitable first approximation is an adiabatic ansatz for the quantum fluctuations,
\begin{align}
    \Delta \hat{A}_\alpha = -c \frac{\omda^2}{N_e\tilde{\omega}_\alpha^2} \tilde\bepsilon_\alpha \cdot \Delta\hatJp \;\text{where}\; \Delta \hat{O} = \hat{O} - \langle \hat{O} \rangle~.
    \label{eq:heisenbergfluctuations}
\end{align}
Surely, also the pure photonic contributions have to be substituted and we obtain
\begin{align}\label{eq:adaggera}
\tilde{\omega}_\alpha \hat{a}_\alpha^\dagger \hat{a}_\alpha \approx \frac{1}{2}\frac{\omda^2}{N_e \tilde{\omega}_\alpha^2} \left(\tilde\bepsilon_\alpha \cdot \hatJp \right)^2
\end{align}
in the adiabatic limit (see App.~\ref{App:breit}). Replacing now all photonic degrees of freedom by their introduced counterparts, we obtain the effective photon-free Hamiltonian
\begin{equation}
\begin{aligned}
    \label{eq:breit}
    &\hat{H}_\mathrm{pf}(t) = -\frac{1}{2} \sum_{i=1}^{N_e} \nabla_i^2+ \sum_{i=1}^{N_e} v(\br_i) + \frac{1}{2}\sum_{i\neq j}^{N_e} w(\br_i,\br_j) \\
    &\;\; +\sum_{\alpha=1}^{M_p} \frac{\tilde{\omega}_\alpha}{2} -\sum_{\alpha=1}^{M_p} \frac{\omda^2}{N_e \tilde{\omega}_\alpha^2} \bigg[ \frac{1}{2}\tilde\bepsilon_\alpha \cdot \left(\hatJp - \Jp(t)\right)\\ &\;\; + \tilde{\omega}_\alpha\int_{-\infty}^t \rmd t' \sin\left(\tilde{\omega}_\alpha(t-t')\right)\tilde\bepsilon_\alpha \cdot \Jp(t') \bigg](\tilde\bepsilon_\alpha \cdot \hatJp)~,
\end{aligned}
\end{equation}
which is now time-dependent.
While computationally simpler to solve, more importantly this provides us with a starting point conceptually much closer to the known realm of electronic structure theory.

A particularly interesting feature of this effective Hamiltonian Eq.~\eqref{eq:breit} is that the adiabatic correction takes the form of a polarization-projected kinetic operator  $-(\tilde\bepsilon_\alpha \cdot \hatJp)^2 \sim (\tilde\bepsilon_\alpha \cdot \nabla_i)^2  + (\tilde\bepsilon_\alpha \cdot \nabla_i)(\tilde\bepsilon_\alpha \cdot \nabla_{j\neq i})$ with opposite sign to the kinetic operator. A large part of the photonic fluctuations are therefore responsible for dynamically increasing the mass of the charged particle along the axis of polarization, very much in line with the perturbative mass-renormalization obtained from the Lamb shift~\cite{craig1998} and the accumulation of electronic density inside cavities~\cite{flick2017c,schafer2019modification,haugland2020intermolecular}. 
If we would extend the number of modes to infinity, we would even recover the logarithmic ultra-violet divergence characteristic of the Lamb shift~\cite{rokaj2020free}. 
We notice furthermore that simply disregarding pure photonic contributions in the substitution procedure from Eq.~\eqref{eq:PF_Hamiltonian} to Eq.~\eqref{eq:breit} would mean to miss the factor $\frac{1}{2}$ introduced in Eq.~\eqref{eq:adaggera}. This would imply that for $\lambda_\alpha \rightarrow \infty$ the mass would not tend to infinity but to negative values instead and the Hamiltonian would become unbounded from below, a common problem for similar approaches within relativistic regimes~\cite{doi:10.1063/1.532834}. 

While we accomplished our initial goal of a non-perturbative photon-free QED theory, such an adiabatic substitution does not come without sacrifices.  Fig.~\ref{fig:spectrumHB} illustrates the absorption spectrum of one-dimensional hydrogen strongly coupled to an optical cavity using the exact Pauli-Fierz Hamiltonian, the photon-free description Eq.~\eqref{eq:breit}, and the purely classical description of the photonic field. Clearly, even on this strongly simplified level including only adiabatic fluctuations, the photon-free description Eq.~\eqref{eq:breit} improves noticeable over the classical Maxwell picture. It accounts correctly for the upwards (1st matter excitation) and down-wards (higher matter excitations) bending at low cavity frequencies.

\begin{figure}
    \centering
    \includegraphics[width=1\columnwidth]{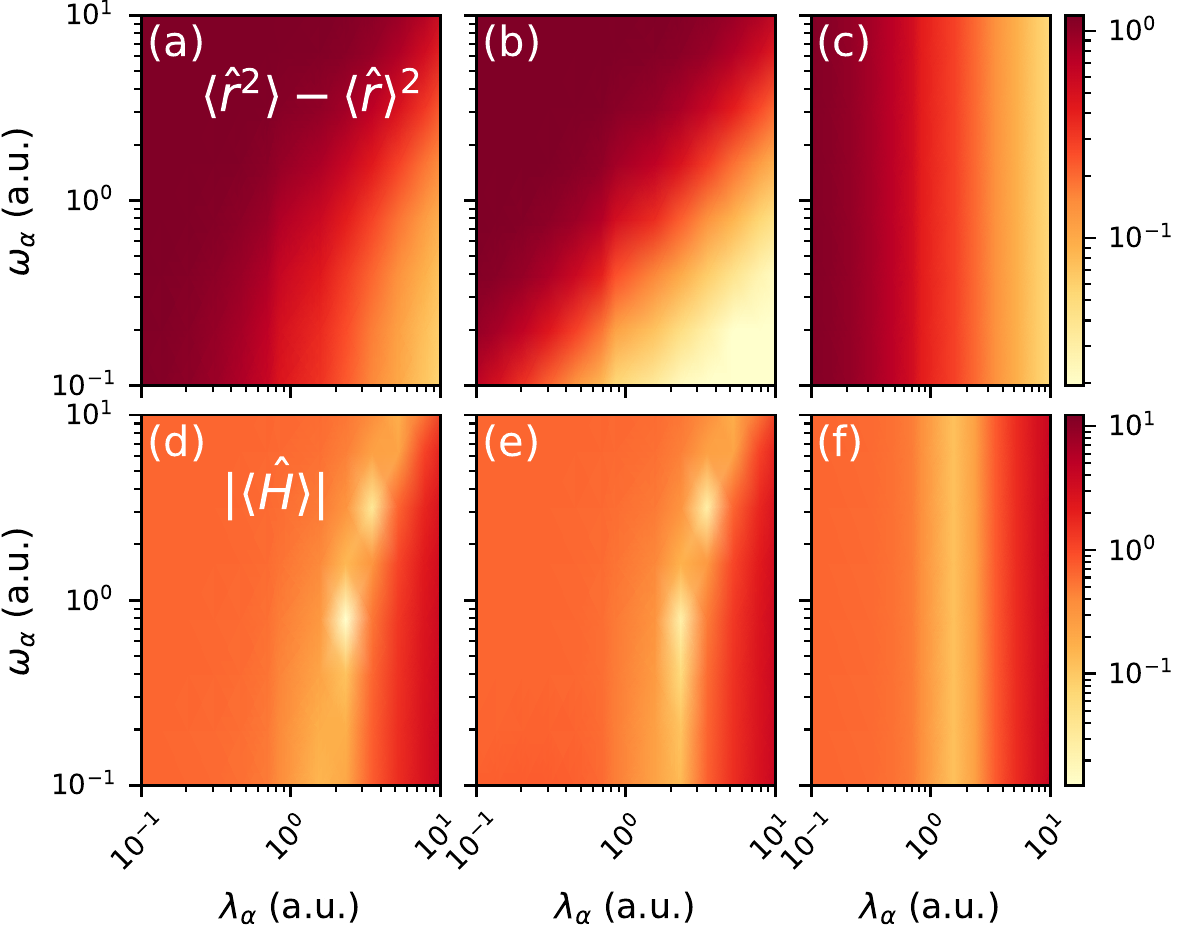}
    \caption{Ground-state variance (a-c) and energy (d-f) for the one-dimensional soft-Coulomb hydrogen atom coupled to a single cavity mode. Illustrated are the exact solution (a,d), the photon-free Hamiltonian Eq.~\eqref{eq:breit} (b,e) and the photon-free self-polarization solution of Eq.~\eqref{eq:PZW_Hamiltonian} (c,f). 
    The results presented here are entirely due to the quantized nature of light, a semi-classical Maxwell treatment would show no effect in the ground state. We disregarded the trivial zero-point energy shift $\omega_\alpha/2$ in all calculations, used 2001 grid points with 0.025$~a_0$ spacing, finite-difference order 4 and 40 photonic Fock states (using the PZW Hamiltonian).}
    \label{fig:benchmarkEBR}
\end{figure}

A particularly important feature of the photon-free Hamiltonian is the explicit dependence on the cavity frequency via ${\omda^2}/{\tilde{\omega}_\alpha^2}$. To understand the importance of this observation, let us utilize the Power-Zienau-Woolley (PZW) transformation~\cite{power1959coulomb,andrews2018perspective,schafer2019relevance} and rewrite our original Coulomb-gauge Hamiltonian Eq.~\eqref{eq:PF_Hamiltonian} in the routinely utilized PZW form \cite{cohen1997photons,craig1998,schafer2019relevance} 
\begin{align}\label{eq:PZW_Hamiltonian}
    \hat{H}_\mathrm{PZW} &= -\frac{1}{2} \sum_{i=1}^{N_e} \nabla_i^2 + \sum_{i=1}^{N_e} v(\br_i) + \frac{1}{2}\sum_{i\neq j}^{N_e} w(\br_i,\br_j)\\ &+ \frac{1}{2}\sum_{\alpha=1}^{M_p} \bigg[ (-\imagi\partial_{p_\alpha})^2 + \omega_\alpha^2 \bigg( p_\alpha + \frac{\lambda_\alpha}{\omega_\alpha}\tilde\bepsilon_\alpha \cdot \sum_{i=1}^{N_e}\br_i \bigg)^2 \bigg]~.\notag
\end{align}
The photonic harmonic-oscillator coordinates are given here explicitly as  $p_\alpha$.
Without truncation of electronic or photonic space, both the Pauli-Fierz Eq.~\eqref{eq:PF_Hamiltonian} and the PZW-Hamiltonian Eq.~\eqref{eq:PZW_Hamiltonian} provide the exact same result for all physical observables, as we would expect from the concept of gauge invariance. We note, however, that the PZW Hamiltonian is not a convenient starting point for extended periodic systems, since the periodicity in the matter coordinates is explicitly broken~\cite{rokaj2017}. Eq.~\eqref{eq:PZW_Hamiltonian} now involves the dipole self-polarization term $\frac{1}{2}(\lambda_\alpha \tilde\bepsilon_\alpha\cdot \sum_{i}\br_i )^2$. So even if we would disregard the photonic degrees of freedom altogether, we would remain with a confining harmonic potential acting on the electronic degrees of freedom that results in the correct behavior for zero and infinite coupling. This approximation is sometimes referred to as QED Hartree-Fock \cite{haugland2020coupled,haugland2020intermolecular}. Fig.~\ref{fig:benchmarkEBR} however clearly illustrates that the photon-free Hamiltonian Eq.~\eqref{eq:breit} is superior due to its explicit frequency dependence. The latter furthermore guarantees that photon-free and exact description coincide for infinite cavity frequency, where matter excitations are no longer allowed to couple to the photonic degrees of freedom. It provides thus an excellent electronic subspace solution which becomes exact whenever a factorized wave-function ansatz is possible. 

By handling operators as if we would treat expectation values, we intrinsically assume that both sides of Eq.~\eqref{eq:heisenbergfluctuations} possess the same set of eigenstates. This, for instance, is satisfied for a set of plane waves describing free electrons and quantized photonic fields. Unfortunately, for the vast majority of condensed matter systems this condition will not be met.
Knowing about the limitation of such a construction, we can however use the solution of the homogeneous electron gas inside the cavity as a basis for a formally more accurate description. We show in the following that even by restricting ourselves to the original electronic Hilbert-space this provides accurate results which become exact in the limit of weak and deep ultra-strong coupling, thus providing a preferable starting point for functional development.

\section{The photon-coupled homogeneous electron gas basis}\label{sec:pheg}

We already stated that the photon-free ansatz becomes exact in the homogeneous limit. This is indeed no coincidence, as it can be shown that in the free ($v=0$) and non-interacting ($w=0$) limit, the Pauli-Fierz Hamiltonian Eq.~\eqref{eq:PF_Hamiltonian} can be diagonalized analytically with a combination of Bogoliubov and coherent shift transformations~\cite{rokaj2020free}. The obtained photon-coupled homogeneous electron gas (pHEG) Hamiltonian is purely additive in the new transformed operators,
\begin{align}\label{eq:pHEG_Hamiltonian}
    \hat{H}_\mathrm{pHEG} = -\frac{1}{2} \sum_{i=1}^{N_e} \nabla_i^2 - \sum_{\alpha=1}^{M_p}\tilde\omega_\alpha \hat\beta_\alpha^2 + \sum_{\alpha=1}^{M_p} \tilde\omega_\alpha \left(\hat{c}_\alpha^\dagger \hat{c}_\alpha + \frac{1}{2}\right)~,
\end{align}
with the coherent shift operator
\begin{align*}
    \hat\beta_\alpha = \frac{\omda}{\sqrt{2N_e\tilde\omega_\alpha^{3}}}  \tilde\bepsilon_\alpha\cdot \sum_{i=1}^{N_e}(-\imagi\nabla_i)~.
\end{align*}
See App.~\ref{App:photon-pHEG} for a definition of the transformed annihilation and creation operators $\hat{c}_\alpha, \hat{c}_\alpha^\dagger$.
An immediate consequence of the additive structure is that the eigenfunctions of the Hamiltonian Eq.~\eqref{eq:pHEG_Hamiltonian} are factorized in nature. Its eigenbasis can be chosen as plane-wave Slater determinants and displaced photon number states,
\begin{align*}
    \vert \{\bk_j\}, \{\beta_\alpha(\bK),n_\alpha\} \rangle =
    \vert \Phi_{\{\bk_j\}} \rangle \otimes \prod_{\alpha=1}^{M_p} \vert \beta_\alpha(\bK),n_\alpha \rangle~.
\end{align*}
Here $\bK = \sum_{i} \bk_j$ represents the collective momentum of all particles, a well defined quantum number in the pHEG system. The photonic states $\vert \beta_\alpha(\bK),n_\alpha \rangle$ implicitly account for the collective momenta of the homogeneous electronic system. It becomes apparent that this new basis, although factorized, represents intrinsically interacting light and matter parts, very similar to the photon-free ansatz motivated in Sec.~\ref{sec:ansatz}. It is now our intention to bring this particular pHEG solution into use as a basis for a general, inhomogeneous system. 
This demands for an expression of the local potential in the pHEG basis, a computationally simple task for any potential that can be represented in a Fourier-basis. The matrix elements of the external single-particle potentials in the pHEG basis then read
\begin{align}\label{eq:V-matrix-el}
    &\langle \{\bk_j\}, \{\beta_\alpha(\bK),n_\alpha\} \vert
    v(\hat\br_i)
    \vert \{\bk_j'\}, \{\beta_\alpha(\bK'),n_\alpha'\} \rangle \\
    &=
    \langle \Phi_{\{\bk_j\}} \vert v(\hat\br_i) \vert \Phi_{\{\bk_j'\}} \rangle \cdot \prod_{\alpha=1}^{M_p} \langle \beta_\alpha(\bK),n_\alpha \vert \beta_\alpha(\bK'),n_\alpha' \rangle~,\notag
\end{align}
where all $\bk_j = \bk'_j$ have to match for $j\neq i$ in order to provide a non-zero expression due to orthogonality.
The first part is the Fourier transformation of the potential at $\bk_i-\bk'_i$ and the second can be evaluated analytically in terms of associated Laguerre polynomials~\cite{CahillGlauber1969} with argument $ \beta_\alpha(\bk_i) - \beta_\alpha(\bk'_i) $. 
We denote the second part as the Fourier transform of a function $m_\alpha^{n,n'}(\br_i)$ and consequently Eq.~\eqref{eq:V-matrix-el} is conveniently expressed with the help of the convolution theorem,
\begin{align*}
    \Fourier v \cdot \prod_{\alpha=1}^{M_p} \Fourier m_\alpha^{n,n'} 
    = \Fourier \left(v * \Conv_{\alpha=1}^{M_p} m_\alpha^{n,n'} \right)~.
\end{align*}
The coherent shifts of the pHEG basis therefore provide an effective screening or mollification of the local potential. In the homogeneous system ($v=0$) photonic excitations $n_\alpha$ are mere replica without substantial relevance. The moment our system becomes inhomogeneous ($v \neq 0$) however, we start to couple different eigenstates via scatterings at the local potential. In perfect agreement with our conclusions in Sec.~\ref{sec:ansatz}, we observe that for an inhomogeneous system the photonic Fock space recovers its relevance.
However, even in lowest order $n=n'=0$, thus entirely within the zero-excitations sector, the modifier is non-zero
\begin{align*}
    m_\alpha^{0,0}(\br) = \sqrt{\frac{N_e\tilde\omega_\alpha^{3}}{\pi\omda^2}} \rme^{-\frac{N_e\tilde\omega_\alpha^{3}}{\omda^2}(\tilde\bepsilon_\alpha\cdot\br)^2} \delta(\tilde\bepsilon_{\alpha}^{\perp,1} \cdot \br) \, \delta(\tilde\bepsilon_{\alpha}^{\perp,2} \cdot \br)~,
\end{align*}
where $\tilde\bepsilon_{\alpha}^{\perp,1},\tilde\bepsilon_{\alpha}^{\perp,2}$ are selected such that they complete $\tilde\bepsilon_{\alpha}$ into an orthonormal basis for $\mathbb{R}^3$.
Convolution with the Gaussian above acts as a mollification and yields a zeroth-order correction to our original photon-free ansatz Eq.~\eqref{eq:breit}. This correction vanishes, i.e., $m_\alpha^{0,0} \rightarrow \delta$, for $\lambda_\alpha \rightarrow 0$ or $\infty$ and $\omega_\alpha \rightarrow \infty$ and we observe again that the photon-free ansatz is exact in those limits. For higher orders $n,n'$ the respectively modified potentials $v * m_\alpha^{n,n'}$ couple different excitation-number sectors.

\begin{figure}
    \centering
    \includegraphics[width=1\columnwidth]{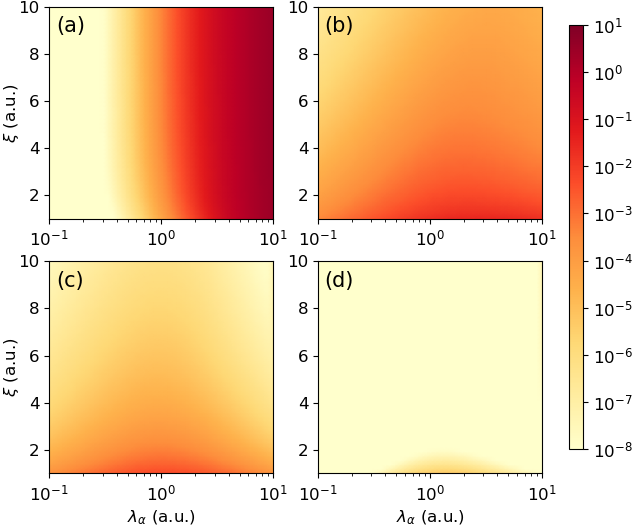}
    \caption{
    Absolute deviation in ground-state energy compared to the exact Pauli-Fierz reference solution of a tunable one-dimensional soft-Coulomb hydrogen  $v(x)=-1/\sqrt{x^2+\xi^2}$
    coupled to a single cavity mode. (a) Pauli-Fierz Hamiltonian with $\max n_\alpha=4$ excitations, (b) pHEG basis with $\max n_\alpha=0$ and the original potential $v$, (c) pHEG basis with $\max n_\alpha=0$ but mollified potential $v*m^{0,0}_\alpha$, (d) pHEG basis with $\max n_\alpha=4$. The pHEG basis is vastly superior over the non-interacting Pauli-Fierz basis for large $\lambda_\alpha$. The flatter the local potential ($\xi \rightarrow \infty$), the closer we get to the scattering-free homogeneous solution and the better performs the pHEG basis.
    The electronic dimension was represented with 31 $k$-points on a periodic grid. The reference solution was obtained using the Pauli-Fierz Hamiltonian with  100 photonic excitations.
    }
    \label{fig:benchmark_cosine_basis}
\end{figure}

Assuming a sufficient number of photonic excitations $n_\alpha$ are considered, the pHEG basis provides exactly the same results as the original Pauli-Fierz Hamiltonian Eq.~\eqref{eq:PF_Hamiltonian}. As presented in Fig.~\ref{fig:benchmark_cosine_basis}, this Fock-space dimension $n_\alpha$ can however be substantially smaller in the pHEG basis while still obtaining much better converged energies than with the trivial non-interacting basis $ \vert \Phi_{\{\bk_j\}}\rangle\otimes \vert n_\alpha \rangle $ for the Pauli-Fierz Hamiltonian. Especially in the ultra-strong coupling limit the superiority of the pHEG basis is apparent. Here, the non-interacting basis demands a quickly growing Fock space while the pHEG basis is exact for $\lambda_\alpha \rightarrow \infty$ by construction, even in its lowest approximation.
Very intuitively, the flatter the local potential, i.e., the closer our system resembles a homogeneous system, the better the pHEG basis converges. 
The sharper the local potential, the more scattering events have to be described by the pHEG basis. Comparing Fig.~\ref{fig:benchmark_cosine_basis} (a) to (d) it however becomes apparent that even for strongly localizing potentials this does not break the ansatz introduced here.
Note that with the change of operators
photon observables are still accessible. Even in the photon-free case matter-fluctuations can represent photonic operators (see, e.g., App.~\ref{App:breit}). While we illustrated here the correlated energy, also purely photonic observables such as the photon number can be accurately reproduced by the truncated pHEG basis with very few excitations, as demonstrated in App.~\ref{App:photon-pHEG}.

In spite of the great advantage that the pHEG basis might provide, the high complexity of the electronic system remains. So even our non-perturbative photon-free ansatz will be challenged by any many-particle system and thus calls for further considerations if we intend to describe realistic molecules or solids. A combination of the photon-free ansatz with established electronic-structure theory approaches will already provide a first suitable description of (ultra-) strong light-matter coupling. The following section illustrates how our previous considerations seamlessly integrate into QEDFT and shows how the conceptions behind the non-perturbative photon-free ansatz can be projected to realistic systems.

\section{Quantum electrodynamical density-functional theory}\label{sec:functionals}

A much more reduced reformulation of QED is QEDFT, where the wave function of the coupled light-matter system is substituted by a current density and a vector potential~\cite{ruggenthaler2011b,tokatly2013,ruggenthaler2014, ruggenthaler2015,jestadt2018real}. Like in other versions of density-functional theory no information is lost by this substitution and the full wave function can be reconstructed in principle from this pair of reduced observables~\cite{ruggenthaler2014}. The main drawback of QEDFT is, similar to any density-functional reformulation, that we in general do not have explicit expressions (expressed only by the pair of reduced observables) for the terms that appear in the defining equations (see, e.g., Eq.~\eqref{eq:Mxcpotential} below). To approximate these terms, one usually relies on auxiliary systems which are easier to treat numerically. The standard choice, that is also followed by and large in QEDFT, is to use non-interacting auxiliary systems. This Kohn-Sham construction gives rise to effective potentials that force the auxiliary non-interacting system to give the same current density and vector potential as the original reference system~\cite{ruggenthaler2014, tokatly2013, ruggenthaler2015,jestadt2018real}. These effective potentials can be expressed by differences of equations of motion. While it is relatively straightforward to find simple approximations for the longitudinal interaction~\cite{doi:10.1146/annurev-physchem-040214-121420,tchenkoue2019force}, for the transverse interactions between light and matter there are only a few approximations hitherto available~\cite{pellegrini2015,flick2017c,flick2021simple}. The main problem with the matter-photon interaction terms in QEDFT is that the auxiliary, uncoupled Kohn-Sham wave function provides an inconvenient starting point to approximate a linearly coupled term of photonic and matter operators (see also the discussion of Eq.~\eqref{eq:Mxcpotential} below). It is at this point where QEDFT can strongly benefit from the methods devised here, and we can make use of the approximate theories derived above.   

Focusing on the ground-state case in the long-wavelength limit, the two basic equations of motion that we will use in order to define the effective potential are the balance of forces due to the paramagnetic current density $\hatjp(\br) = \frac{1}{2\imagi}\sum_{i} \left(\delta(\br -\br_i)\overrightarrow{\nabla_i} - \overleftarrow{\nabla_i}\delta(\br -\br_i)\right)$,
\begin{align}\label{eq:EOMpara}
 \rho(\br)\nabla v(\br) = \langle\hat\FF_T(\br)\rangle + \langle\hat\FF_W(\br)\rangle -\frac{1}{c}\left\langle \big(\hat{\bA} \cdot \nabla\big) \hatjp(\br) \right\rangle~,
\end{align}
together with the static mode-resolved Maxwell's equations,
\begin{align}\label{eq:MaxwellQEDFT}
  A_{\alpha} =  -\frac{c \omda^2}{N_e \tilde{\omega}_{\alpha}^2} \tilde{\bepsilon}_{\alpha}\cdot \Jp~.
\end{align}
Both equations follow from the Heisenberg equation of motion for their respective operators (see also Eq.~\eqref{eq:heisenberg}) with the Pauli-Fierz Hamiltonian of Eq.~\eqref{eq:PF_Hamiltonian} and are evaluated with the ground state of the coupled light-matter system $\Psi$. Here the density operator is $\hat{\rho}(\br) =  \sum_{i} \delta(\br_i-\br)$, $\hat{\FF}_T(\br) = \frac{\imagi}{2} [\hat{\bj}_{\rm p}(\br), \sum_{i} \nabla^2_{r_{i}}]$ is the local stress force, $\hat{\FF}_W(\br) = -\frac{\imagi}{2} [\hat{\bj}_{\rm p}(\br), \sum_{i\neq j} w(\br_i,\br_j)]$ is the local interaction force, and $\langle \cdot \rangle$ indicates the expectation value with respect to $\Psi$. Alternatively, Eqs.~\eqref{eq:EOMpara} and \eqref{eq:MaxwellQEDFT} can be derived from the corresponding full minimal-coupling expression (see App.~\ref{App:mcMxapproximation}) by taking the long-wavelength limit. In the static case the zero-component of the current density, i.e., the density $\rho(\br)$, determines also the other components of the current density~\cite{ruggenthaler2015}. Thus the exact mean-field exchange-correlation (Mxc) scalar potential of static Kohn-Sham QEDFT is defined by
\begin{align}\label{eq:Mxcpotential}
    &\nabla^2 v_{\rm Mxc}(\br) = \nabla\cdot\frac{1}{\rho(\br)}\Big[\FF_T([\Phi],\br) - \FF_T([\Psi],\br) \\
    &- \FF_W([\Psi],\br) +\frac{1}{c}\left\langle \big(\hat{\bA} \cdot \nabla\big) \hat{\bj}_{\rm p}(\br)\right\rangle - \frac{1}{c}(\bA \cdot \nabla) \bj_{\rm p}([\Phi],\br) \Big], \nonumber
\end{align}
where $\FF_T([\Phi],\br)$ and $\bj_{\rm p}([\Phi],\br)$ indicate the expectation values of $\hat{\FF}_T(\br)$ and $\hat{\bj}_{\rm p}(\br)$ with respect to the non-interacting Kohn-Sham wave function $\Phi$, and $\FF_T([\Psi],\br)$ as well as $\FF_W([\Psi],\br)$ accordingly with respect to $\Psi$. Further, in the static case the mean-field contribution $(\bA \cdot \nabla) \bj_{\rm p}[\Phi]$, which arises from the $\hat{\bJ}_{\rm p}\cdot \bA$ coupling in the Kohn-Sham system, is zero. For the differences in stress and interaction forces various approximations based on the Kohn-Sham wave function exist in the DFT literature~\cite{tokatly2005quantum,tokatly2005,B903666K,PhysRevLett.79.4878}. Taking just the non-interacting Kohn-Sham wave function already leads to a non-vanishing expression for the interaction force $\FF_{W}([\Phi],\br)$. This contribution is called the local Hartree-exchange contribution~\cite{PhysRevA.80.052502,tchenkoue2019force},
\begin{align}\label{eq:Hxpotential}
    \nabla^2 v_{\rm Hx}(\br) = -\nabla \cdot \left[\frac{\FF_W([\Phi], \br) }{\rho(\br)}\right].
\end{align}
The same is no longer the case for the matter-photon interaction term in Eq.~\eqref{eq:Mxcpotential}, where substituting just the auxiliary Kohn-Sham wave function leads to zero. The reason for this is that we work with an uncoupled and non-interacting Kohn-Sham wave function (the photonic part only consists of trivial shifted harmonic oscillators~\cite{ruggenthaler2015, jestadt2018real}). Having an approximation to the coupling in terms of matter quantities becomes therefore highly desirable.

The most straightforward approach follows the discussion in Sec.~\ref{sec:ansatz} by just replacing $\Delta \hat{A}_{\alpha} = \hat{A}_{\alpha} - A_{\alpha} \rightarrow -c \omda^2/(N_e \tilde{\omega}^2_{\alpha}) \tilde{\bepsilon}_{\alpha}\cdot \Delta \hatJp$ in Eq.~\eqref{eq:Mxcpotential}. In order to guarantee the real-valuedness of the expectation value we have to use the symmetrized form of Eq.~\eqref{eq:forceBreit}. While on the level of the equations of motion the validity of this \textit{ad hoc} ansatz remains unclear, we find that by using the photon-free Hamiltonian of Eq.~\eqref{eq:breit} to derive the equation of the paramagnetic current density the approximate coupling term becomes
\begin{align} \label{eq:forceBreit}
  &\frac{1}{c}\left\langle \big(\Delta \hat{\bA} \cdot \nabla\big) \hatjp(\br)\right\rangle \\
  &\longrightarrow -\frac{1}{2}\sum_{\alpha=1}^{M_p} \frac{\omda^2}{N_e \tilde{\omega}_{\alpha}^2}\left( \left\langle \big(\tilde{\bepsilon}_{\alpha}\cdot\Delta\hatJp\big) \big(\tilde{\bepsilon}_{\alpha}\cdot\nabla\big) \hatjp(\br)\right\rangle + c.c. \right),\nonumber
\end{align}
which exactly agrees with the above symmetrized equation-of-motion-based substitution. Here the expectation value is taken with respect to the photon-free QED ground-state wave function of Eq.~\eqref{eq:breit}. From this alternative approach that leads to the same expressions we see that this simple approximation becomes exact for the various limiting cases discussed in the previous sections. We furthermore know how to constructively increase the accuracy of this approach by including more terms from the pHEG basis expansion with the respective mollification. This implies that it is a reasonable strategy to build approximations for the coupling-force term based on the simple substitution of Eq.~\eqref{eq:forceBreit}. Rewriting the exact Mxc potential as $v_{\rm Mxc}(\br) = v_{\rm px}(\br) + v_{\rm Hx}(\br) + v_{\rm c}(\br)$, we find the static orbital-dependent photon-exchange (px) contribution as
\begin{align}
    \label{eq:staticHxpotential}
    \nabla^2 v_{\rm px}(\br) &= -\nabla \cdot \left[\sum_{\alpha=1}^{M_p} \frac{\omda^2}{N_e \tilde{\omega}_{\alpha}^2} \frac{ (\tilde{\bepsilon}_{\alpha}\cdot\nabla) \left\langle \left( \tilde{\bepsilon}_{\alpha} \cdot \hat{\bJ}_{\rm p} \right) \hat{\bj}_{\rm p}(\br) \right\rangle}{\rho(\br)}\right], 
\end{align}
where now the expectation value is taken with respect to a real-valued, auxiliary Kohn-Sham wave function and thus also $\bJ_{\rm p} \equiv 0$. This expression can be further simplified for special cases, such as in one spatial dimension and for one particle ($\Psi=\varphi=\sqrt{\rho}$), where it becomes 
\begin{align}\label{eq:meanfieldexchange1D}
    \partial_x^2 v_{\rm px}(x)\! &=\! -\!\!\sum_{\alpha=1}^{M_p} \frac{\omda^2}{2 \tilde{\omega}^2_{\alpha}} \partial_x\!\!\left[\frac{\partial_x \!\left[\left( \partial_x\varphi(x)\right)^2 \! - \! \left( \partial^2_x \varphi(x)\right)  \! \varphi(x)\right]}{\varphi(x)^2}\right] \notag \\
    &=\sum_{\alpha=1}^{M_p} \frac{\omda^2}{2 \tilde{\omega}^2_{\alpha}} \partial_x^2\left[\frac{\partial^2_x \sqrt{\rho(x)} }{\sqrt{\rho(x)}}\right].
\end{align}
Eq.~\eqref{eq:meanfieldexchange1D} intuitively illustrates that electronic density becomes accumulated at local maxima, consistent with increasing the particle mass along the polarization direction as discussed in Sec.~\ref{sec:ansatz}.

However, in general the expression for $v_{\rm px}$ (Eq.~\eqref{eq:staticHxpotential}) is still an orbital-dependent functional and hence can become costly for very large systems. It is therefore desirable to simplify this expression even further. Borrowing from a recent demonstration that Eq.~\eqref{eq:Hxpotential} leads to the well-known exchange-only local-density approximation (LDA) for the Coulomb interaction in the homogeneous case~\cite{tchenkoue2019force}, we follow the same strategy to devise a simple LDA for Eq.~\eqref{eq:staticHxpotential}. 

Starting point of the derivation, which we provide in detail in App.~\ref{App:Mx-LDA}, is to express the current-current correlation in terms of one and two-body reduced density matrices
\begin{align*}
    &\left\langle \left( \tilde{\bepsilon}_{\alpha} \cdot \hat{\bJ}_{\rm p} \right) \hat{\bj}_{\rm p}(\br) \right\rangle \\
    &=\frac{1}{2} \left[ (\tilde{\bepsilon}_{\alpha} \cdot \nabla') \nabla \rho_{(1)}(\br,\br') - ( \tilde{\bepsilon}_{\alpha} \cdot \nabla') \nabla' \rho_{(1)}(\br,\br') \right]_{\br'=\br} \\
    &+\int \left[( \tilde{\bepsilon}_{\alpha} \cdot \nabla_2') \nabla \rho_{(2)}(\br,\br_2;\br',\br_2') + c.c.\right]_{\br'=\br,\br_2'=\br_2} \rmd\br_2~.
\end{align*}
From here on, we use the closed-shell exchange representation of the two-body reduced density matrix
$\rho_{(2)}(\br,\br_2;\br',\br_2') = \tfrac{1}{2}[\rho_{(1)}(\br,\br')\rho_{(1)}(\br_2,\br_2') - \tfrac{1}{2} \rho_{(1)}(\br,\br_2')\rho_{(1)}(\br_2,\br')]$
for spin-$\tfrac{1}{2}$ particles
in combination with the homogeneous electron-gas ansatz $ \rho_{(1)}(\br,\br') = 2(2\pi)^{-d} \int_{\vert \bk \vert < \kF } \exp({\imagi\bk\cdot(\br-\br')}) \rmd\bk$ with a local Fermi radius $k_F$ in $d$ spatial dimensions.
%
This leads to the photon-exchange-only LDA (pxLDA) that can be given for arbitrary spatial dimensionality $d$,
\begin{align}\label{eq:lda}
    \nabla^2 v_{\mathrm{pxLDA}}(\br) =  - \sum_{\alpha=1}^{M_p} \frac{2\pi^2\omda^2}{N_e\tilde\omega_\alpha^2} (\tilde{\bepsilon}_{\alpha} \cdot \nabla)^2 \left( \frac{\rho(\br)}{2V_d} \right)^{\!\!\frac{2}{d}}
\end{align}
with $V_d$ the volume of the $d$-dimensional unit sphere.
For a single electron, an additional factor 2 would appear, but here we will further stick to the given closed-shell form (see App.~\ref{App:Mx-LDA} for a more broad discussion).
Let us emphasize that even this particularly simple functional includes the correct frequency and cavity-polarization dependence that are inherent to the non-perturbative photon-free approach. Assuming an isotropic interaction between light and matter, Eq.~\eqref{eq:lda} allows for the direct solution $ v_{\mathrm{pxLDA}}(\br) = -\sum_\alpha 2\pi^2\omda^2 / (d N_e\tilde\omega_\alpha^2) ( \rho(\br)/(2V_d))^{2/d}$, a form suitable to describe free-space Lamb-shift physics. However, this form is consequently inapplicable to cavity settings.

The local potentials from Eqs.~\eqref{eq:Mxcpotential}-\eqref{eq:lda} are generally determined as the solutions to a Poisson-type equation $ \nabla^2 v(\textbf{r}) = -f(\textbf{r})$. Solving the latter is a routinely performed calculation-step in \textit{ab initio} density-functional theory codes in order to obtain the Hartree potential $\nabla^2 v_\mathrm{H}(\textbf{r})=-4\pi\rho(\textbf{r})$ in a cost-efficient way~\cite{tancogne2019octopus}.
It is thus straightforward to go to realistic, three-dimensional systems. The numerical implementation and validation of such functionals for sizable, realistic systems will be discussed in detail in a forthcoming publication.  

Finally, the energy associated with the $v_{\rm Mx} = v_{\rm Hx} + v_{\rm px} $, or approximations thereof, is given by 
\begin{align*}
E_\mathrm{Mx} &= \left\langle -\frac{1}{2} \sum_{i=1}^{N_e} \nabla_i^2 + \frac{1}{2}\sum_{i\neq j}^{N_e} w(\br_i,\br_j) \right\rangle + \int v(\br) \rho(\br) \rmd \br  \\
& +\sum_{\alpha=1}^{M_p} \frac{\omda^2}{2N_e\tilde{\omega}_\alpha^2} \left\langle \sum_{i,j=1}^{N_e}(\tilde{\bepsilon}_\alpha \cdot \nabla_i) (\tilde{\bepsilon}_\alpha \cdot \nabla_{j}) \right\rangle + \sum_{\alpha=1}^{M_p}\frac{\tilde{\omega}_\alpha}{2}
\end{align*}
with the expectation values with respect to the non-interacting Kohn-Sham wave function $\Phi$ determined by $v_{\rm Mx}$.
Further, photonic observables can be approximated by using the simple substitution discussed in Sec.~\ref{sec:ansatz} and also in App.~\ref{App:breit}.

\begin{figure}
    \centering
    \includegraphics[width=1\columnwidth]{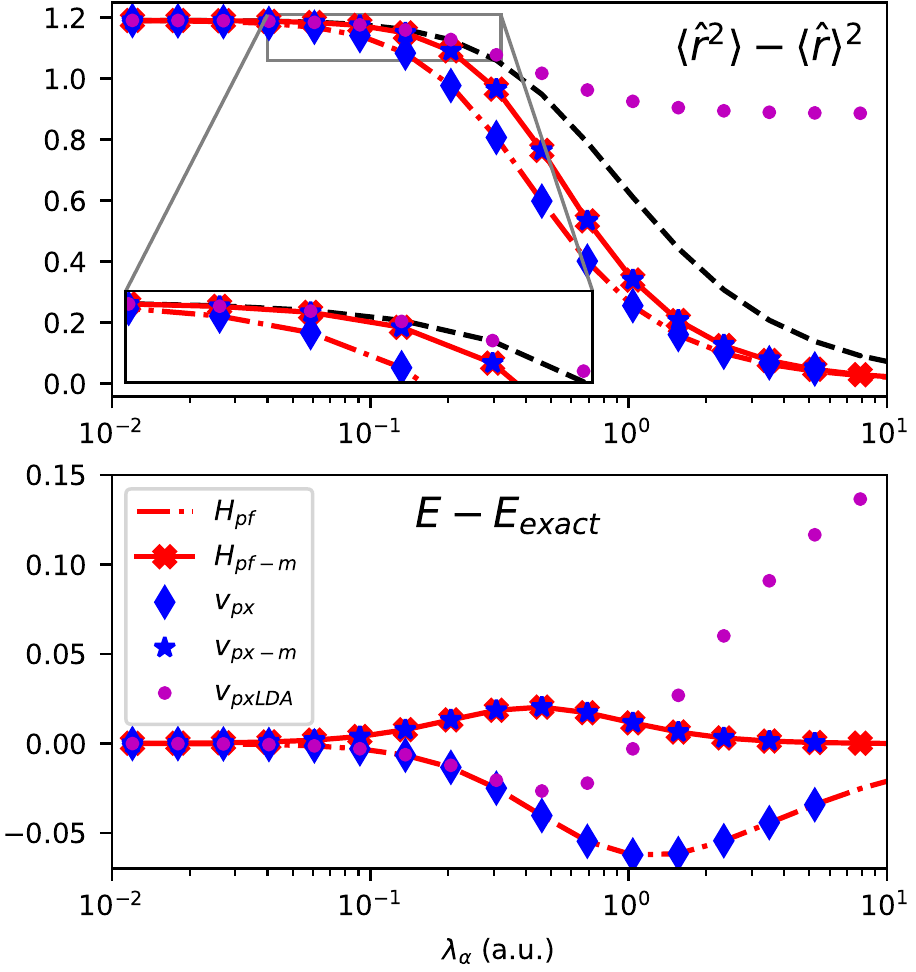}
    \caption{Dipole variance $\langle \hat{r}^2 \rangle - \langle \hat{r} \rangle^2 $ (top) and total energy difference to the exact solution $E-E_{\rm exact}$ (bottom) of the correlated cavity hydrogen system for various approximations compared to the exact solution (black dashed) with 40 photon number-states (using the PZW Hamiltonian). The index ``$-m$'' indicates the utilization of the \textit{ad hoc} mollification of the external potential according to Sec.~\ref{sec:pheg}. The mollification leads to variational energies. The exchange LDA potential provides excellent results up to $\lambda=0.3$.
    The cavity frequency of the single cavity mode is set in resonance with the bare lowest excitation energy like in Fig.~\ref{fig:spectrumHB}. The real-space grid and potential was chosen identical to Fig.~\ref{fig:benchmarkEBR}. Since we selected particularly tough parameters for the benchmark, the approximations presented here will perform better for softer potentials (compare Sec.~\ref{sec:pheg}).}
    \label{fig:benchmark_integrated}
\end{figure}

Let us next consider how these simple exchange-type approximations for the coupling between light and matter perform in a test-case scenario. In Fig.~\ref{fig:benchmark_integrated} we show the spatial electronic dipole variance and total energy for a one-dimensional soft-Coulomb hydrogen model from weak to deep ultra-strong coupling. Investigating a single-particle system guarantees that we only consider the reliability of the light-matter approximation and do not mix in the approximate longitudinal electron-electron description.
By construction, the photon-exchange approximation $v_{\rm Mx}$ (blue diamonds) recovers the photon-free QED Hamiltonian solution (red dashed-dotted line). Both (consistent) approximations over-estimate the coupling effect (the exact result is the black dashed line) and tend to over-bind. As discussed in Sec.~\ref{sec:ansatz}, these approximations become exact for weak and deep ultra-strong coupling. In the intermediate (ultra-strong) regime employing the mollification (crossed-line and blue stars) suggested by the basis-expansion in Sec.~\ref{sec:pheg} leads to a clear improvement. The mollification also renders the minimization variational, i.e., by construction the energy is always above the exact energy. Similarly to the usual Coulombic LDA, which tends to delocalize electrons too strongly, the pxLDA also underestimates the enhanced binding. This leads to a partial error cancellation in the ultra-strong coupling domain but results in the wrong infinite-coupling limit. 

\begin{figure}
    \centering
    \includegraphics[width=1.0\columnwidth]{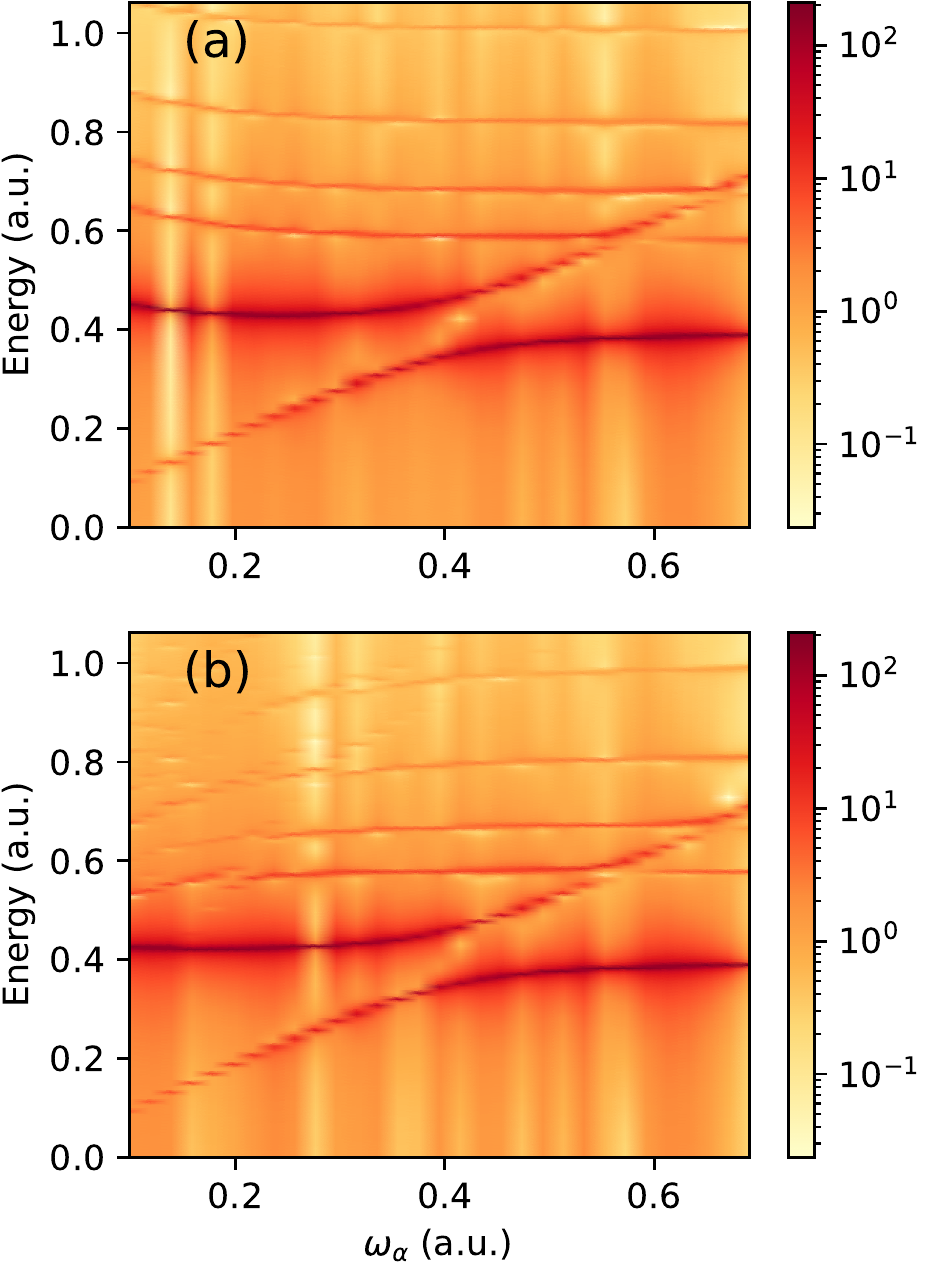}
    \caption{Linear response spectrum of the dipole moment $\br(\omega_\alpha)$ of a single cavity mode with varying frequency $\omega_\alpha$ coupled to one-dimensional soft-Coulomb hydrogen. Shown are the adiabatic utilization of the pxLDA potential Eq.~\eqref{eq:lda} in combination with the semi-classical Maxwell solution (a, top) and the exact reference solution (b, bottom). All parameters are identical to Fig.~\ref{fig:spectrumHB}.}
    \label{fig:spectrumHBLDA}
\end{figure}

Overall, the pxLDA provides fairly accurate predictions for ultra-strong light-matter coupling with minimal additional costs when compared to common density-functional approximations for the Coulombic interaction. This computational simplicity allows us to utilize the pxLDA in an adiabatic manner also for time-propagation. Fig.~\ref{fig:spectrumHBLDA} presents the linear-response spectrum (similar to Fig.~\ref{fig:spectrumHB}) and illustrates that the adiabatic pxLDA potential correctly predicts the upwards-bending of the first matter-excitation at low frequencies. However, it falsely predicts the same behavior for all excited states.
There are several possible origins for that. Firstly, since in the time-dependent case we have complex wave functions, in the original Eq.~\eqref{eq:staticHxpotential} we should work with $\Delta \hat{\bJ}_{\rm p}$ instead of $\hat{\bJ}_{\rm p}$. Thus Eq.~\eqref{eq:lda} also includes the implicit approximation $\bJ_{\rm p} \equiv 0$ due to the adiabatic homogeneous-electron-gas ansatz for the fluctuations. This can be corrected by working with Eq.~\eqref{eq:staticHxpotential} and by re-substituting $\Delta \hat{\bJ}_{\rm p}$. Secondly, in the time-dependent case we should actually use a different equation of motion (total current instead of just the paramagnetic current), which leads to the contribution of several further terms that are zero in the static case, and we should use a consistent approximation to the Mxc vector potential as well (see App.~\ref{App:tdMxapproximation}). That is, to be consistent with the static approximation we should actually use a current-density formulation of QEDFT also in the dipole-coupling limit. 
In addition, we need to include non-adiabatic photon fluctuation effects. This poses a real problem for a simple Hamiltonian approach as discussed in Sec.~\ref{sec:ansatz} since it is inconsistent to mix the operators in the Heisenberg picture with the Hamiltonian in the Schr\"odinger picture used to propagate the wave functions. Yet, for the equation-of-motion approach employed in this section, no such restriction applies, since we can easily switch between the different pictures. This allows to define also a non-adiabatic version of the presented px potential (as discussed in App.~\ref{App:tdMxapproximation}) leading to multi-photon excitations. 

Furthermore, for a detailed understanding of light-matter coupling effects,
it is desirable to go beyond the common dipole approximation. Again we can simply follow the above strategy and replace the fluctuations of the vector-potential operator by inverting the inhomogeneous quantum Maxwell equation, which is discussed in App.~\ref{App:mcMxapproximation}. Yet since we can derive the defining equation for the px potential by taking the long-wavelength limit of the corresponding minimal-coupling equations (see also App.~\ref{App:mcMxapproximation}), we can find a first approximation to the beyond-dipole case by simply making the coupling terms $\omda \propto \lambda^2_{\alpha}$ and the polarizations $\tilde{\bepsilon}_{\alpha}$ spatially dependent in Eq.~\eqref{eq:staticHxpotential} or \eqref{eq:lda}.
Such a simple extension would already lead to (static) ponderomotive forces that allow to trap polarizable systems in the eigenmode profile.

Finally, we note that we have limited our initial investigation
presented here to the exchange-level of theory in the longitudinal and the transverse light-matter interactions. Besides going beyond the single-Slater-determinant ansatz~\cite{PhysRevA.80.052502,fuks2016time} or combinations with wave-function-based methods~\cite{mordovina2019self}, there is the possibility (hitherto only for the longitudinal interaction) to use directly parametrizations or approximations of correlation expressions~\cite{parr-yang-book,PhysRevA.87.022514}. The photon-free framework provides us with functionals that are in close structural relation to the known approximations and conditions developed in ordinary density-functional theory which facilitates the further development of QEDFT.
Providing accurate yet computationally affordable functionals that go beyond the dipole approximation, include correlations and non-adiabatic contributions will be the subject of future work.

\section{Conclusion}

Solving the Schr\"odinger equation for realistic system is an extremely hard task. 
The moment we consider the quantum character of light, we add an abundance of additional degrees of freedom to the already almost intractable problem.
Our goal to retain the first-principles character of electronic-structure theory and entwine it with QED is thus particularly challenging, unless we find effective descriptions which allow us to absorb large parts of the QED problem into the familiar electronic problem.

We presented here such an effective photon-free framework which adheres to all essential physical demands, including the correct frequency and cavity-polarization dependence, and accounts even in its simplest form for a substantial part of the full QED problem. Even if we remain entirely within the electronic Hilbert space, we recover the exact results of light-matter interaction in the weak coupling, deep ultra-strong coupling and high frequency limit, as well as for homogeneous systems. This framework provides an excellent electronic subspace solution which becomes exact whenever a factorized wave-function ansatz is possible.
Starting from there, we illustrated that the photon-free approach corresponds to a highly efficient basis for light-matter interaction in many relevant situations.
Expressing the coupling purely in matter quantities provided a convenient starting point to develop exchange-correlation potentials dealing with correlated light-matter systems.
We then leveraged the potential of the photon-free ansatz by constructing a hierarchy of QEDFT functionals, illustrated their performance for ground and excited states and discussed how to go beyond the simple adiabatic and dipole-approximated cases. As outcome we derived the first local-density approximated functional, combining the quantum nature of light with the computational simplicity of well known electronic local-density functionals in DFT. The photon-free construction, including the derived QEDFT functionals, is ideally suited to describe periodic systems strongly coupled to light.

That the photon-free QED Hamiltonian remains entirely within the electronic Hilbert space is of great benefit not only computationally but also conceptually. The latter feature allows it to serve as a foundation for the development of more advanced electron-photon-structure approaches without the need to consider extensions of the known methodologies to larger Fock spaces. For instance, the description of strong light-matter coupling in terms of non-equilibrium Green's functions commonly introduces the necessity to solve an expanded set of Kadanoff-Baym equations~\cite{mahan2013many}. The present photon-free QED ansatz on the other hand could serve as a starting point that remains exact in the infinite coupling limit. It is thus ideal for perturbative improvements on top of its non-perturbative foundation.

Our approach is conceptually general enough to be expanded to vibrational light-matter coupling and beyond the dipole approximation. Extensions beyond the dipole approximation become feasible if we follow the equation-of-motion constructions of Sec.~\ref{sec:functionals}.
The obtained spectral information and the lack of multi-photon excitations suggests that future development should foremost focus on going beyond the adiabatic approximation. While this demands curing an old wound of density-functional theory, the equation-of-motion construction provides a passage to memory-dependent functionals. It furthermore avoids the numerical and perturbative instability of other orbital-dependent functionals such as the time-dependent optimized-effective potential~\cite{wijewardane2008real}. This would not only greatly benefit the description of strongly-correlated light-matter systems but also ameliorate common problems with ordinary density-functional theory.

\begin{acknowledgments}
We thank G\"oran Johansson for helpful comments. 
This work was supported by the European Research Council (ERC-2015-AdG694097), the Cluster of Excellence ``CUI: Advanced Imaging of Matter'' of the Deutsche Forschungsgemeinschaft (DFG) EXC 2056 (project ID 390715994), Grupos Consolidados (IT1249-19), the Federal Ministry of Education and the Research Grant RouTe-13N14839, the SFB925 ``Light induced dynamics and control of correlated quantum systems'', and the Swedish Research Council (VR) through Grant No. 2016-06059. M.~P.\ acknowledges support by the Erwin Schrödinger Fellowship J 4107-N27 of the FWF (Austrian Science Fund).
\end{acknowledgments}

\input{appendices}

\bibliographystyle{apsrev4-2}
\bibliography{breit}
\end{document}

%% file: appendices.tex
\appendix

\section{The Bogoliubov transformation}\label{App:bogo}

In the following, we recall the well-known Bogoliubov transformation~\cite[Sec.~1.10]{Faisal1987} utilized in Sec.~\ref{sec:ansatz}. The transformation is based on the realization that the purely photonic part of Hamiltonian Eq.~\eqref{eq:PF_Hamiltonian}
\begin{align*}
	\hat{H}_{\rm ph}=&\sum_{\alpha=1}^{M_p} \omega_{\alpha}\left(\hat{a}^{\dagger}_{\alpha}\hat{a}^{\dphan}_{\alpha}+\frac{1}{2}\right) + \frac{N_e}{2 c^2} \hat{\bA}^2
\end{align*}
with $\hat\bA$ given by Eq.~\eqref{eq:A_starting_def} is simply the Hamiltonian of $M_p$ coupled harmonic oscillators. This is even more obvious when we introduce the harmonic oscillator coordinates $q_{\alpha}=1/\sqrt{2\omega_{\alpha}} \left(\hat{a}^{\dagger}_{\alpha}+\hat{a}_{\alpha}\right)$ and $p_{\alpha}=\imagi\sqrt{\omega_{\alpha}/2} \left(\hat{a}^{\dagger}_{\alpha}-\hat{a}_{\alpha}\right)$, which leads to the Hamiltonian
\begin{align*}
	\hat{H}_{\rm ph}=&\frac{1}{2}\left(\sum_{\alpha=1}^{M_p} p_{\alpha}^2 + \smashoperator{\sum_{\alpha,\alpha'=1}^{M_p}} W_{\alpha,\alpha'}q_{\alpha} q_{\alpha'}\right) \label{eq:coupledHO},
\end{align*}
with $W_{\alpha,\alpha'}=\omega_{\alpha}^2 \delta_{\alpha,\alpha'} + 4\pi N_e S_{\alpha}(\br_0)S_{\alpha'}(\br_0)\bm{\bepsilon}_{\alpha}\cdot\bm{\bepsilon}_{\alpha'}$. Since $W_{\alpha,\alpha'}$ is symmetric there is a unitary transformation $U$ which brings $W_{\alpha,\alpha'}$ into diagonal form $\tilde{\Omega}=UWU^{\dagger}$ with eigenvalues $\tilde{\omega}_\alpha^2$. This diagonalization introduces the decoupled normal modes with corresponding operators $\tilde{q}_{\beta}=\sum_{\alpha} U_{\beta,\alpha}q_{\alpha}$, $\tilde{p}_{\beta}=\sum_{\alpha} U_{\beta,\alpha}p_{\alpha}$ and polarization vectors $\tilde{\bm{\epsilon}}_{\beta}=\sum_{\alpha} U_{\beta,\alpha}\bm{\epsilon}_{\alpha}$.
The resulting Hamilton reads
\begin{align*}
    \hat{H}_{\rm PF} &= -\frac{1}{2} \sum_{i=1}^{N_e} \nabla_i^2 + \sum_{i=1}^{N_e} v(\textbf{r}_i) + \frac{1}{2}\sum_{i\neq j}^{N_e} w(\textbf{r}_i,\textbf{r}_j) \\ &+ \frac{1}{c} \hatJp \cdot \hat{\bA} + \sum_{\alpha=1}^{M_p} \tilde{\omega}_\alpha \left(\hat{\tilde{a}}_\alpha^\dagger \hat{\tilde{a}}_\alpha + \frac{1}{2}\right),
\end{align*}
where we introduced the annihilation operators $\hat{\tilde a}_\alpha=1/\sqrt{2 \tilde{\omega}_\alpha}(\tilde{\omega}_\alpha \tilde{q}_{\alpha} + \imagi\tilde{p}_{\alpha})$ corresponding to $\tilde{q}_{\alpha}, \tilde{p}_{\alpha}$, as well as the respective creation operators.

\section{Adiabatic Breit approximation for the photon energy}\label{App:breit}

A sensible effective photon-free Hamiltonian (e.g., Eq.~\eqref{eq:breit}) has to adhere to the same fundamental rules as the original Hamiltonian. Foremost, this includes the existence of bound eigenstates and translational invariance.
To achieve this, we cannot just disregard all photonic degrees of freedom in an \textit{ad hoc} fashion but instead have to find the matter term corresponding to $\sum_{\alpha} \tilde{\omega}_\alpha \hat{a}_\alpha^\dagger \hat{a}_\alpha$. From the equations of motion we get for the annihilation operators (the creation operators are always just the Hermitian conjugate),
\begin{equation*}
\frac{\rmd^2}{\rmd t^2} \hat{a}_{\alpha}
	=-\tilde{\omega}_{\alpha}^2 \hat{a}_{\alpha} - \frac{\omda\sqrt{\tilde{\omega}_{\alpha}}}{\sqrt{2N_e}} \bm{\bepsilon}_{\alpha} \cdot \left( \hatJp - \frac{\imagi}{\tilde{\omega}_{\alpha}}  \frac{\rmd}{\rmd t} \hatJp \right)~.
\end{equation*}
we can identify the adiabatic approximation as
\begin{equation*}
    \hat{a}_{\alpha}\approx - \frac{\omda}{\sqrt{2N_e\tilde{\omega}_{\alpha}^3}} \bm{\bepsilon}_{\alpha} \cdot \hatJp~.
\end{equation*}
This leads to the critical contribution that renders the photon-free Hamiltonian bounded from below
\begin{equation*}
     \tilde{\omega}_\alpha \hat{a}_\alpha^\dagger \hat{a}_\alpha \approx \frac{\omda^2}{2N_e\tilde{\omega}_{\alpha}^2} (\bm{\bepsilon}_{\alpha} \cdot \hatJp)^2~.
\end{equation*}

\section{Photon observables in the truncated pHEG basis}
\label{App:photon-pHEG}

The Bogoliubov and coherent shift transformations of $\hat{a}_\alpha,\hat{a}_\alpha^\dagger$ mentioned in Sec.~\ref{sec:pheg} can be combined as
\begin{equation*}
	\hat{c}_\alpha =  \sqrt{\frac{\tilde\omega_\alpha}{4\omega_\alpha}} \left(\hat{a}_\alpha^\dagger + \hat{a}_\alpha\right) - \sqrt{\frac{\omega_\alpha}{4\tilde\omega_\alpha}} \left(\hat{a}_\alpha^\dagger - \hat{a}_\alpha\right) + \hat\beta_\alpha~,
\end{equation*}
with the back transformation
\begin{equation*}
    \hat{a}_\alpha = \frac{1}{2}\left( \sqrt{\frac{\omega_\alpha}{\tilde\omega_\alpha}}\left(\hat{c}_\alpha^\dagger+\hat{c}_\alpha\right) - \sqrt{\frac{\tilde\omega_\alpha}{\omega_\alpha}}\left(\hat{c}_\alpha^\dagger-\hat{c}_\alpha\right) \right) - \sqrt{\frac{\omega_\alpha}{\tilde\omega_\alpha}}\hat\beta_\alpha~.
\end{equation*}
In a photon-free approach that ignores excited states of $\hat{c}_\alpha^\dagger \hat{c}_\alpha$ the photon number operator for mode $\alpha$ is thus found to be $\hat{a}_\alpha^\dagger \hat{a}_\alpha = \frac{\omega_\alpha}{\tilde\omega_\alpha}\hat\beta_\alpha^2$. In the limits $\omega_\alpha \to 0$ and $\omega_\alpha \to \infty$ this operator goes to zero as expected. The expectation value of the photon number operator for the pHEG ground state with $\beta_\alpha(\bK)=0$ and $n_\alpha=0$ is found to be $\langle\hat{a}_\alpha^\dagger \hat{a}_\alpha\rangle = (\tilde\omega_\alpha-\omega_\alpha)^2/(4\tilde\omega_\alpha\omega_\alpha)$ \cite{rokaj2020free} (the factor two in the reference comes from taking two different polarization directions into account).

\begin{figure}
    \centering
    \includegraphics[width=1\columnwidth]{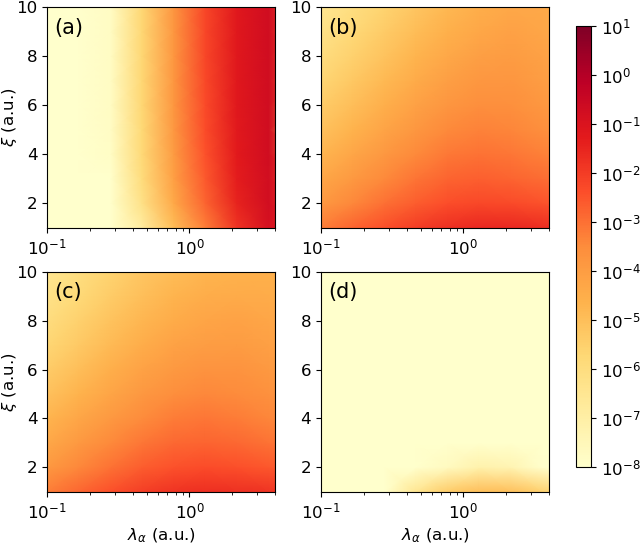}
    \caption{
    Absolute deviation of the photon number in the ground state compared to a Pauli-Fierz reference solution for a tunable one-dimensional soft-Coulomb hydrogen $v(x)=-1/\sqrt{x^2+\xi^2}$
    coupled to a single cavity mode. (a) Pauli-Fierz Hamiltonian with $\max n_\alpha=4$ excitations, (b) pHEG basis with $\max n_\alpha=0$ and the original potential $v$, (c) pHEG basis with $\max n_\alpha=0$ but mollified potential $v*m^{0,0}_\alpha$, (d) pHEG basis with $\max n_\alpha=4$. The result is similar to Fig.~\ref{fig:benchmark_cosine_basis}, just with a lesser benefit from the mollified potential.
    The electronic dimension has 41 $k$-points on a periodic grid. The reference solution was obtained using the Pauli-Fierz Hamiltonian with 100 photonic excitations.
    The displayed parameter area is smaller than in Fig.~\ref{fig:benchmark_cosine_basis} in order to limit it to values where the reference solution has converged.
    }
    \label{fig:benchmark_cosine_basis_photoncount}
\end{figure}

\begin{figure}
    \centering
    \includegraphics[width=1\columnwidth]{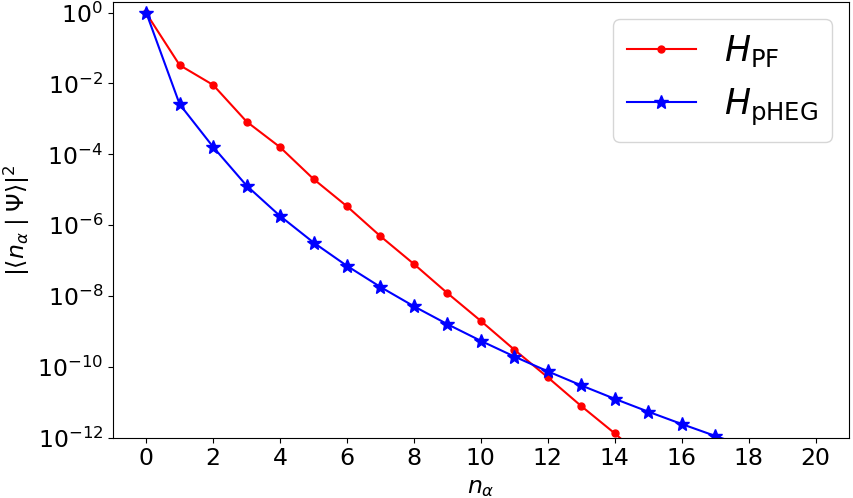}
    \caption{
    Excitation-number distribution for the ground state with a pHEG basis compared to a Pauli-Fierz reference solution for a one-dimensional soft-Coulomb potential with $\xi=1$ coupled to a single cavity mode with $\lambda_\alpha=1$.
    The electronic dimension has 41 $k$-points on a periodic grid. The photon filling in the low-number sectors in considerably reduced in the pHEG basis which makes the truncation at low excitation numbers numerically accurate.
    }
    \label{fig:benchmark_cosine_basis_photonfilling}
\end{figure}

A benchmark calculation is displayed in Fig.~\ref{fig:benchmark_cosine_basis_photoncount} and shows a rapid increase in accuracy including just a few excitation numbers in a truncated pHEG basis. This special feature of the pHEG approximation is further highlighted by plotting the excitation-number distribution, i.e., the probability to find the system in each excitation-number sector, see Fig.~\ref{fig:benchmark_cosine_basis_photonfilling}. For small excitation numbers the pHEG basis shows a much quicker decrease, meaning a higher accuracy if the basis is truncated at low excitation numbers.

\section{Photon-exchange-only LDA derivation}
\label{App:Mx-LDA}

\begin{figure}
    \centering
    \includegraphics[width=1\columnwidth]{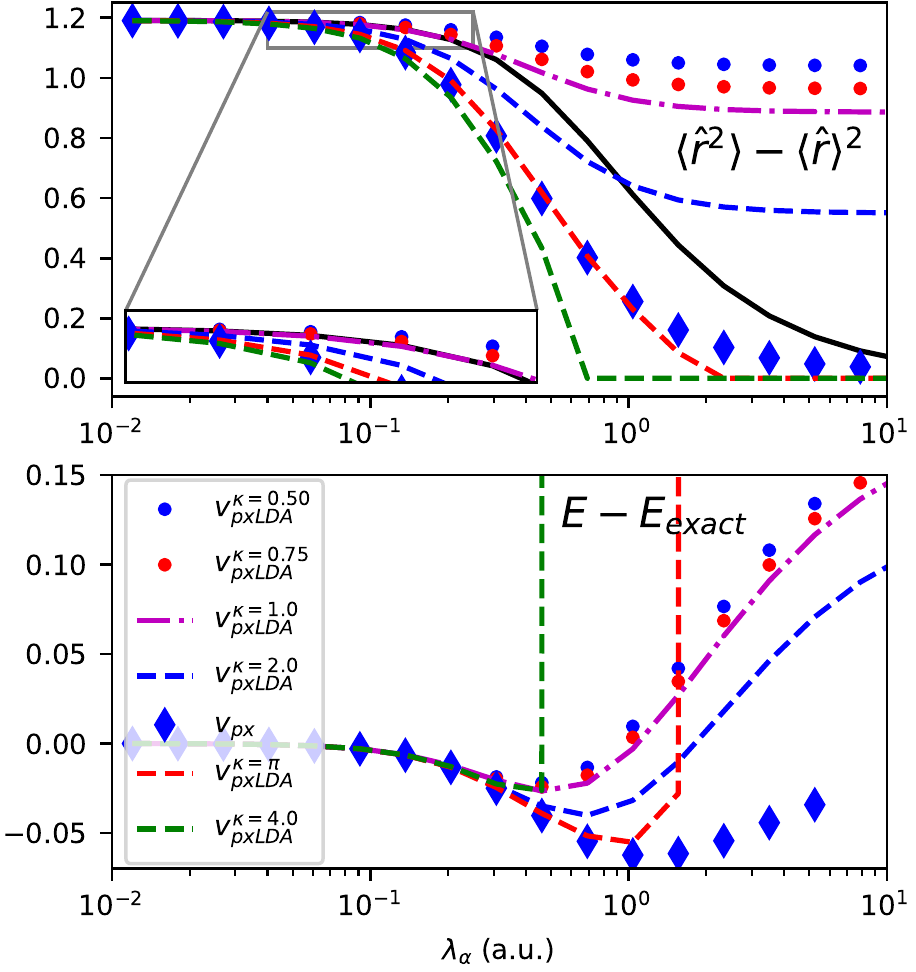}
    \caption{
    Exchange LDA approximation $v_{\rm pxLDA}^{\kappa}$ with different parameters $\kappa$ compared to the exact reference solution (black solid) for the one-dimensional soft-Coulomb potential with $\xi=1$ coupled to a single cavity mode in resonance to the first bare excitation energy. The setup is identical to Fig.~\ref{fig:benchmark_integrated}.
    The choice $\kappa=1$ suggested here provides excellent results up to $\lambda=0.3$. Larger values can improve the (deep) ultra-strong coupling limit at the cost of instability (collapse onto a few/single grid-point) and overestimation in the strong to ultra-strong domain.
    }
    \label{fig:benchmark_LDAkappa}
\end{figure}

In order to arrive at a local-density approximation for the photon-exchange-only term, we note that for a properly anti-symmetrized wave function $\Phi(\br,\obr)$, $\obr=(\br_2\ldots\br_{N_e})$, we can express the expectation value in Eq.~\eqref{eq:staticHxpotential} in terms of the one-body (1RDM) and two-body reduced density matrices (2RDM)~\cite[Sec.~2.4]{parr-yang-book} as
\begin{equation}\label{eq:app-LDA:f-RDM}
\begin{aligned}
    &\mathbf{f}_\alpha(\br) = \left\langle \left( \tilde{\bepsilon}_{\alpha} \cdot \hat{\bJ}_{\rm p} \right) \hat{\bj}_{\rm p}(\br) \right\rangle = \left\langle \left( \tilde{\bepsilon}_{\alpha} \cdot \hat{\bJ}_{\rm p} \right) \Phi,\, \hat{\bj}_{\rm p}(\br) \Phi \right\rangle \\
    &=\frac{N_e}{2\imagi} \int \left[ \left( \left( \tilde{\bepsilon}_{\alpha} \cdot \hat{\bJ}_{\rm p} \right)\Phi\right)^* \nabla \Phi - \left( \left( \tilde{\bepsilon}_{\alpha} \cdot \hat{\bJ}_{\rm p} \right) \nabla\Phi\right)^* \Phi \right] \rmd\obr \\
    &=\frac{N_e}{2} \int \left[ ((\tilde{\bepsilon}_{\alpha} \cdot \nabla)\Phi^*) \nabla \Phi - ((\tilde{\bepsilon}_{\alpha} \cdot \nabla)\nabla\Phi^*) \Phi \right] \rmd\obr \\
    &+\frac{N_e(N_e-1)}{2} \int \underbrace{\left[ ((\tilde{\bepsilon}_{\alpha} \cdot \nabla_2)\Phi^*) \nabla \Phi - ((\tilde{\bepsilon}_{\alpha} \cdot \nabla_2)\nabla\Phi^*) \Phi \right]}_{((\tilde{\bepsilon}_{\alpha} \cdot \nabla_2)\Phi^*) \nabla \Phi + c.c.} \rmd\obr \\
    &=\frac{1}{2} \left[ (\tilde{\bepsilon}_{\alpha} \cdot \nabla') \nabla \rho_{(1)}(\br,\br') - ( \tilde{\bepsilon}_{\alpha} \cdot \nabla') \nabla' \rho_{(1)}(\br,\br') \right]_{\br'=\br} \\
    &+\int \left[( \tilde{\bepsilon}_{\alpha} \cdot \nabla_2') \nabla \rho_{(2)}(\br,\br_2;\br',\br_2') + c.c.\right]_{\br'=\br,\br_2'=\br_2} \rmd\br_2~.
\end{aligned}
\end{equation}
We used partial integration to move $\tilde{\bepsilon}_{\alpha} \cdot \nabla_2$ to the other side in the underbraced expression. This 2RDM contribution will later be found having the same form as the 1RDM terms but with opposite sign, lessening the strength of the attractive px potential, and will finally be neglected altogether.
In the remaining integral the 2RDM can be rewritten in terms of the 1RDM for closed-shell Slater-determinant states of spin-$\tfrac{1}{2}$ particles that we assume from here, $\rho_{(2)}(\br,\br_2;\br',\br_2') = \tfrac{1}{2}[\rho_{(1)}(\br,\br')\rho_{(1)}(\br_2,\br_2') - \tfrac{1}{2} \rho_{(1)}(\br,\br_2')\rho_{(1)}(\br_2,\br')]$. This makes the integral
\begin{equation}\label{eq:app-LDA:2RDM-part}
\begin{aligned}
    \int (\tilde{\bepsilon}_{\alpha} \cdot \nabla_2') &\nabla \rho_{(2)}(\br,\br_2;\br',\br_2')\big\vert_{\br'=\br,\br_2'=\br_2} \rmd\br_2 \\
    = \frac{1}{2} \int &\Big[ \nabla \rho_{(1)}(\br,\br')\big\vert_{\br'=\br} (\tilde{\bepsilon}_{\alpha} \cdot \nabla_2') \rho_{(1)}(\br_2,\br_2')\big\vert_{\br_2'=\br_2} \\
    &- \frac{1}{2} ((\tilde{\bepsilon}_{\alpha} \cdot \nabla_2)\nabla \rho_{(1)}(\br,\br_2)) \rho_{(1)}(\br_2,\br) \Big] \rmd\br_2 \\
    = \frac{1}{2} \nabla & \rho_{(1)}(\br,\br')\big\vert_{\br'=\br} \int (\tilde{\bepsilon}_{\alpha} \cdot \nabla_2') \rho_{(1)}(\br_2,\br_2')\big\vert_{\br_2'=\br_2} \rmd\br_2 \\
    &- \frac{1}{4} \int ((\tilde{\bepsilon}_{\alpha} \cdot \nabla_2)\nabla \rho_{(1)}(\br,\br_2)) \rho_{(1)}(\br_2,\br) \rmd\br_2~. \notag
\end{aligned}
\end{equation}
For the homogeneous electron gas of spin-$\tfrac{1}{2}$ particles the 1RDM is then given by the usual Fermi-sphere integration~\cite[Sec.~6.1]{parr-yang-book}
\begin{equation}\label{eq:app-LDA:1RDM-HEG}
    \rho_{(1)}(\br,\br') = \frac{2}{(2\pi)^d} \smashoperator{\int_{\vert \bk \vert < \kF }} \rme^{\imagi\bk\cdot(\br-\br')} \rmd\bk~,
\end{equation}
with the local Fermi radius given by $\kF(\br) = 2\pi (\rho(\br)/(2V_d))^{1/d}$ with $V_d$ the volume of the $d$-dimensional unit sphere (see \citet[Eq.~(1.77)]{giuliani-vignale-book}, but also easily derived from \eqref{eq:app-LDA:1RDM-HEG} by setting $\br'=\br$). The Fermi sphere is centered around the origin since we consider the static case with zero current ${\bj}_{\rm p}=0$. Then  
\begin{equation*}
    \nabla \rho_{(1)}(\br,\br')\big\vert_{\br'=\br} = \frac{2\imagi}{(2\pi)^d} \smashoperator{\int_{\vert \bk \vert < \kF }} \bk\, \rmd\bk = 0
\end{equation*}
is readily seen to be zero because $\bk$ is integrated over the symmetric Fermi-sphere volume. We now insert the ansatz \eqref{eq:app-LDA:1RDM-HEG} also into all remaining terms from \eqref{eq:app-LDA:f-RDM} and get with $\kF=\kF(\br)$, chosen at $\br$ because this is the primary position, and $\kF'$ left open that
\begin{equation*}
\begin{aligned}
    &\mathbf{f}_\alpha(\br) = \frac{2}{(2\pi)^d} \smashoperator{\int_{\vert \bk \vert < \kF }} (\tilde{\bepsilon}_{\alpha} \cdot \bk)\bk \rmd\bk \\
    &- \frac{1}{(2\pi)^{2d}} \smashoperator{\iint_{\substack{\vert \bk \vert < \kF\\ \vert \bk' \vert < \kF' }}} (\tilde{\bepsilon}_{\alpha} \cdot \bk)\bk \bigg(\underbrace{\int
     \rme^{\imagi\bk\cdot(\br-\br_2)} \rme^{\imagi\bk'\cdot(\br_2-\br)} \rmd\br_2}_{\rme^{\imagi (\bk-\bk')\cdot\br} (2\pi)^d\delta(\bk-\bk')} + c.c.\bigg) \rmd\bk \rmd\bk' \\
     &= \frac{2}{(2\pi)^d}  \smashoperator{\int_{\vert \bk \vert < \kF }} (\tilde{\bepsilon}_{\alpha} \cdot \bk)\bk \rmd\bk - \frac{2}{(2\pi)^d}  \smashoperator{\int_{\vert \bk \vert < \min(\kF,\kF') }} (\tilde{\bepsilon}_{\alpha} \cdot \bk)\bk \rmd\bk~.
\end{aligned}
\end{equation*}
Here the $\br_2$ integration led to a delta function that leaves the smaller integration radius from the two $\bk,\bk'$ integrals. What was done here is to introduce the local approximation Eq.~\eqref{eq:app-LDA:1RDM-HEG} for $\rho_{(1)}(\br,\br_2)$ even though the positions $\br,\br_2$ are not close. Since further $\br_2$ is taken from the whole space, we can argue that for an inhomogeneous medium $\min(\kF,\kF')$ approaches zero because $k_F'$ possibly gets very small and thus no contribution arises from the original 2RDM expression.
On the other hand, for a homogeneous medium it will hold $\min(\kF,\kF')=\kF$ since $k_F'=k_F$ and thus $\mathbf{f}_\alpha(\br)=0$ which exactly fits our expectations.
To take all those different situations into account, we introduce a factor $\kappa \in [0,1]$ that expresses the ratio that is left from the first integral after subtracting the second with a smaller radius. The maximally inhomogeneous limit corresponds to $\kappa=1$ while the homogeneous case is $\kappa=0$. 
Within the main text we decided to limit ourselves entirely to the case $\kappa=1$. Considering spin-polarized systems motivates a spin-resolved LDA which would suggest a different $\kappa$ regime and will be the subject of future work. Introducing the new parameter $\kappa$ is a simple approach that allows us to stay in the realm of the local density approximation. 
Clearly, those considerations suggest that the light-matter interaction, as it is non-local in character, should be described ideally by more advanced non-local functionals following the spirit of modern (meta) GGAs. Fig.~\ref{fig:benchmark_LDAkappa} illustrates the performance of the LDA for various values also including the possibility $\kappa>1$. In this example, the LDA potential with $\kappa=1$ provides even better results than the full exchange potential for $\lambda < 0.3$ due to error-compensation.

Evaluating the integral by switching to polar coordinates we get
\begin{equation*}
    \mathbf{f}_\alpha(\br) = \frac{2\kappa}{(2\pi)^d}  \smashoperator{\int_{\vert \bk \vert < \kF }} (\tilde{\bepsilon}_{\alpha} \cdot \bk)\bk \rmd\bk = \frac{2\kappa V_d}{(2\pi)^d} \frac{\kF(\br)^{d+2}}{d+2} \tilde{\bepsilon}_{\alpha}~.
\end{equation*}
%
Putting this solution for $\mathbf{f}_\alpha(\br)$ into Eq.~\eqref{eq:staticHxpotential} yields a Poisson-like equation for the photon-exchange-only LDA potential,
\begin{align*}
    &\nabla^2 v_{\rm pxLDA}(\br) = -\nabla \cdot \left[ \sum_{\alpha=1}^{M_p} \frac{\omda^2}{N_e\tilde\omega_\alpha^2} \frac{(\tilde{\bepsilon}_{\alpha} \cdot \nabla)\mathbf{f}_\alpha(\br)}{\rho(\br)} \right] \\
    &= -\frac{2\kappa V_d}{(2\pi)^d} \sum_{\alpha=1}^{M_p} \frac{\omda^2}{N_e\tilde\omega_\alpha^2} (\tilde{\bepsilon}_{\alpha} \cdot \nabla) \left[ \frac{\kF(\br)^{d+1}}{\rho(\br)} (\tilde{\bepsilon}_{\alpha} \cdot \nabla) \kF(\br) \right].
\end{align*}
Expressing the Fermi radius by the local density once more, we get the final result
\begin{equation*}
    \nabla^2 v_{\rm pxLDA}(\br) = - \sum_{\alpha=1}^{M_p} \frac{2\kappa\pi^2\omda^2}{N_e\tilde\omega_\alpha^2} (\tilde{\bepsilon}_{\alpha} \cdot \nabla)^2 \left( \frac{\rho(\br)}{2V_d} \right)^{\!\!\frac{2}{d}}.
\end{equation*}
Assuming an isotropic mode distribution, the differential operators $(\tilde{\bepsilon}_{\alpha} \cdot \nabla)^2$ are summed up to full Laplacians and thus, if zero boundary conditions are assumed, the photon-exchange-only LDA potential is explicitly given by
\begin{equation*}
\begin{aligned}
    v_{\rm pxLDA}(\br) &= - \sum_{\alpha=1}^{M_p} \frac{2\kappa\pi^2\omda^2}{d N_e\tilde\omega_\alpha^2} \left( \frac{\rho(\br)}{2V_d} \right)^{\!\!\frac{2}{d}}
\end{aligned}
\end{equation*}

\section{Minimal-coupling photon-exchange approximation}
\label{App:mcMxapproximation}

The static Pauli-Fierz Hamiltonian in full minimal coupling in SI units takes the form~\cite{spohn2004,ruggenthaler2014}
\begin{align*}
\hat{H}(t) =& \sum_{i=1}^{N_e}\frac{1}{2m} \left( -\imagi \hbar \nabla_i - \frac{q}{c}\hat{\bA}(\br_i) \right)^2 + \sum_{i=1}^{N_e} \underbrace{q a_{0}(\br_i)}_{=v(\br_i)} \\ 
-& \sum_{i=1}^{N_e} \frac{q}{2mc}\boldsymbol{\sigma}_i\cdot\left(\nabla_i \times \hat{\bA}(\br_i)\right) + \frac{1}{4 \pi \epsilon_0}\sum_{i<j}^{N_e}\frac{q^2}{|\br_i-\br_j|} \\
+&  \sum_{s=1}^{2}\int \hbar \omega(\bk) \hat{a}^{\dagger}(\bk,s) \hat{a}(\bk,s) \rmd \bk~,
\end{align*}
where additionally an external current can be coupled to the photon subsystem to establish the basic mapping theorem of ground-state QEDFT~\cite{ruggenthaler2014, jestadt2018real}. The vector potential operator in Coulomb gauge is given by 
\begin{align*}
    \hat{\bA}(\br) =  \sqrt{ \frac{\hbar c^2}{\epsilon_{0} (2 \pi)^3 } }
          \int \frac{\rmd \bk}{\sqrt{2 \omega_{k}}} 
          \sum \limits_{s=1}^2 &\bepsilon(\bk,s) 
          \left[ 
          \hat{a}(\bk,s)\mathrm{e}^{ \imagi \bk \cdot \br} \right.
          \\
          &\left. +\, \hat{a}^{\dagger}(\bk,s) \rme^{-\imagi \bk \cdot \br}
          \right].
\end{align*}
We use the usual definitions for the frequency $\omega(\bk) = c|\bk|$, the bosonic creation and annihilation field operators $\hat{a}^{\dagger}(\bk,s)$ and $\hat{a}(\bk,s)$, as well as for the polarization unit vectors $\bepsilon(\bk,s)$ for the continuum of free-space modes indicated by their wave vector $\bk$ and the two physical polarization directions $s$. Further, $a^{0}(\br)$ is a scalar external vector potential and $\boldsymbol{\sigma}$ denotes a vector of the $2\times2$ Pauli matrices in the Stern-Gerlach term. The Pauli-Fierz Hamiltonian can be derived either by taking the non-relativistic limit of the Dirac Hamiltonian~\cite{ruggenthaler2014}, where the Stern-Gerlach term appears naturally, or by quantizing the Abraham model of classical radiation-reactions with an \textit{ad hoc} inclusion of the Stern-Gerlach term~\cite{spohn2004}. In order for the Pauli-Fierz Hamiltonian to be well-defined we need to include an ultra-violet cutoff and we further note that $m$ is the bare mass of the electrons. The main reason for working in SI units in this appendix is to keep track of the difference between bare and physical mass.

Following the discussion in the main text we will use two basic equations of motion to establish the minimal-coupling px approximation. The first one is the operator form of the Maxwell's equation in the Heisenberg picture~\cite{ruggenthaler2014,jestadt2018real}
\begin{align}
\label{eq:Maxwelloperator}
&  \left(\frac{1}{c^2}\frac{\rmd}{\rmd t} -\nabla^2 \right) \hat{\bA}_{\rm H}(\br,t) = \mu_{0} c q \; \hat{\bj}_{\perp, \rm H}(\br,t) ,
\end{align}
where the physical-current-density operator in the Schr\"odinger picture is
\begin{equation}
        \begin{alignedat}{2}     
             \hat{\bj}(\br)
          &= \hat{\bj}_{\rm p}(\br) + \hat{\bj}_{\rm d}(\br) + \hat{\bj}_{\rm m}(\br). \notag
        \end{alignedat}   
\end{equation}
The additional $\perp$ indicates that we only consider the divergence-free part due to the Coulomb gauge~\cite{ruggenthaler2014, jestadt2018real} and the first term in the total current density is the paramagnetic current density
      \begin{align}
           \hat{\bj}_{\rm p}(\br)
         = \frac{\hbar}{2 m \imagi} \sum_{i=1}^{N_e}
             \left( \delta(\br -\br_i)\overrightarrow{\nabla_i} - \overleftarrow{\nabla_i}\delta(\br -\br_i) \right)~, \notag
      \end{align}
      the second term is the diamagnetic current density
      \begin{align}
          \hat{\bj}_{\rm d}(\br)
        = - \frac{q}{m c}
          \sum_{i=1}^{N_e} \delta(\br-\br_i) \hat{\bA}(\br)~, \notag
      \end{align}
      and the last one is the magnetization current due to the Stern-Gerlach term
      \begin{align}
          \hat{\bj}_{\rm m}(\br) 
        =  \sum_{i=1}^{N_e} \frac{\hbar}{2 m}\left(\delta(\br-\br_i)  \overrightarrow{\nabla_i} \times \boldsymbol{\sigma}_i +   \overleftarrow{\nabla_i} \times \boldsymbol{\sigma}_i \delta(\br-\br_i)\right)~. \notag
      \end{align}

The second equation of motion, which will be enough to investigate the static case in analogy to the main text, is the (component wise) paramagnetic equation of motion~\cite{ruggenthaler2014, jestadt2018real}
\begin{align}\label{eq:MinCoupPara}
    &\frac{\rmd}{\rmd t} \hat{j}^{k}_{\rm p,H}(\br,t) = \hat{F}^{k}_{T,{\rm H}}(\br,t) + \hat{F}^k_{W,{\rm H}}(\br,t) \\
    & + \frac{q}{mc} \sum_{l=1}^3 \left[ \hat{A}^{l}_{\rm H}(\br,t) \partial_l \hat{j}^k_{\rm p,H}(\br,t) + \left(\partial_k \hat{A}^{l}_{\rm H}(\br,t)\right) \hat{j}^{l}_{\rm p,H}(\br,t)  \right] \nonumber \\
    &-\frac{1}{m}\left[ \partial_k \left(\frac{q^2}{2 m c^2}\hat{\bA}_{\rm H}(\br,t)^2 + v(\br)  \right)\right]\hat{\rho}_{\rm H}(\br,t) \nonumber \\
    &+ \frac{q}{ mc} \sum_{l,m,n =1}^{3}\left(\partial_k \partial_l \hat{A}^{m}_{\rm H}(\br,t)  \right) \epsilon^{lmn} \hat{\mu}_{\rm H}^{n}(\br,t)~, \nonumber
\end{align}
where $\epsilon^{lmn}$ is the anti-symmetric Levi-Civita symbol and
\begin{align*}
    &\hat{\FF}_{T}(\br) = \frac{\imagi \hbar^2}{2 m} \left[\hat{\bj}_{\rm p}(\br),\sum_{i} \nabla^2_i\right] ~,\\
    &\hat{\FF}_{W}(\br) = -\frac{\imagi}{4 \pi \epsilon_0} \left[\hat{\bj}_{\rm p}(\br),\sum_{i < j}^{N_e} \frac{q^2}{|\br_i-\br_j|} \right] ~,\\
    &\hat{\boldsymbol{\mu}}(\br) = \frac{\hbar}{2 m}\sum_{i} \boldsymbol{\sigma}_{i} \delta(\br-\br_i) ~.
\end{align*}
Using now a Pauli-Kohn-Sham system with the exact $\bA(\br) = \langle \hat{\bA}(\br)\rangle $ from the Maxwell-Kohn-Sham equation~\cite{jestadt2018real} we get the same Eq.~\eqref{eq:MinCoupPara}, where instead of $\hat{\bA}(\br)$ we just use the expectation value (mean-field) $\bA(\br)$ and replace $v(\br)$ by $v_{\rm s}(\br)$. Using now that $v_{\rm Mxc}(\br) = v_{\rm s}(\br) - v(\br)$ we find (suppressing the $\br$ dependency)
\begin{widetext}
\begin{equation}\label{eq:MinCoupMHxc}
\begin{aligned}
    &\nabla^2 v_{\rm Mxc} = m \sum_{k=1}^{3} \partial_k \frac{1}{\rho} \bigg[ F^{k}_{T}[\Phi] - F^{k}_{T}[\Psi] - F^{k}_{W}[\Psi]  - \frac{q}{mc} \sum_{l=1}^3 \left(\langle \hat{A}^{l} \partial_l \hat{j}^k_{\rm p}\rangle - A^l \partial_l j^{k}_{\rm p}[\Phi] + \left\langle \left(\partial_k \hat{A}^{l} \right)  \hat{j}^{l}_{\rm p} \right\rangle - \left(\partial_k A^{l} \right) j^{l}_{\rm p}[\Phi]   \right)  \\
    &+ \frac{1}{m}\left(  \left\langle \left(\partial_k \frac{q^2}{2 m c^2}\hat{\bA}^2 \right) \hat{\rho} \right\rangle  - \left(\partial_k \frac{q^2}{2 m c^2} \bA^2 \right) \rho \right) 
    -\frac{q}{ mc} \sum_{l,m,n =1}^{3} \left(\left\langle \left(\partial_k \partial_l \hat{A}^{m}  \right) \epsilon^{lmn} \hat{\mu}^n \right\rangle - \left(\partial_k \partial_l A^{m}  \right) \epsilon^{lmn} \mu^n[\Phi]  \right)\bigg]~.
\end{aligned}
\end{equation}
\end{widetext}
Again, the major issue is to find a reasonable approximation for the explicit light-matter coupling terms. Following the discussion in the main text we solve Eq.~\eqref{eq:Maxwelloperator} formally, consider the fluctuations about the mean-field $\Delta \hat{\bA}(\br)$ and replace
\begin{align}\label{eq:MinCoupAnsatz}
    \Delta \hat{\bA}(\br) \rightarrow \frac{q}{4 \pi \epsilon_0 c }\int \frac{\Delta \hat{\bj}_{\perp}(\br')}{|\br -\br'|} \rmd \br' 
\end{align}
in Eq.~\eqref{eq:MinCoupMHxc} (in a symmetrized manner as discussed in the main text) and use $\Psi \rightarrow \Phi$ throughout. In this way, we have defined the corresponding minimal-coupling px approximation, where the usual quantum-mechanical Hx part is defined by Eq.~\eqref{eq:Hxpotential}.

Assuming now that the induced fields have wavelengths that are much larger than the extension of the matter subsystem of interest, we can make the long-wavelength approximation for the px potential. Considering Eq.~\eqref{eq:MinCoupMHxc}, the only term that is not strongly suppressed in this case is
\begin{align}
   &\frac{q}{mc} \sum_{l=1}^3\left(\left\langle \hat{A}^{l} \partial_l \hat{j}^k_{\rm p}\right\rangle -  A^l \partial_l j^{k}_{\rm p}[\Phi] \right) \notag\\
   &\longrightarrow \frac{q}{mc} \left(\left\langle \big(\hat{\bA}\cdot\nabla\big) \hat{\bj}_{\rm p} \right\rangle - (\bA\cdot\nabla) \bj_{\rm p}[\Phi]\right)~. \nonumber
\end{align}
Since in the long-wavelength limit also the Stern-Gerlach term vanishes, the physical current is just $\hat{\bj}(\br) = \hat{\bj}_{\rm p}(\br) + \hat{\bj}_{\rm d}(\br)$. Furthermore, replacing the Green's function of the free-space Laplacian by its periodic finite-volume counterpart 
\begin{align}\label{eq:CoulombModes}
     \frac{1}{4 \pi |\br-\br'|} \rightarrow \sum_{\boldsymbol{n} \in \mathbb{Z}^3}\frac{1}{V\bk_{\boldsymbol{n}}^2} \rme^{\imagi \bk_{\boldsymbol{n}}\cdot (\br-\br')}~,
\end{align}
where $\bk_{\boldsymbol{n}} = \tfrac{2 \pi}{L} \boldsymbol{n}$ and $V=L^3$, we can express Eq.~\eqref{eq:MinCoupAnsatz} explicitly by
\begin{align*}
    \Delta \hat{\bA}(\br) = \frac{q}{\epsilon_0 c}\sum_{\boldsymbol{n}, s} \frac{\bepsilon_{\boldsymbol{n},s}}{V\bk_{\boldsymbol{n}}^2} \int \rme^{\imagi \bk_{\boldsymbol{n}}\cdot (\br-\br')} \bepsilon_{\boldsymbol{n},s}\cdot \Delta \hat{\bj}(\br') \rmd \br'~.
\end{align*}
If we now denote $\alpha \equiv (\boldsymbol{n},s)$, $S_{\alpha}(\br) = \exp(\imagi \bk_{\boldsymbol{n}}\cdot \br)/\sqrt{V}$, $\lambda_{\alpha}(\br) =  S_{\alpha}(\br)\sqrt{1/\epsilon_0}$ and $\omega_{\rm d}^2 = q^2 N/(m \epsilon_0 V)$ this becomes in the long-wavelength limit
\begin{align}\label{eq:varlimit}
\Delta \hat{\bA} = \frac{c}{q N}\sum_{\alpha} \bepsilon_{\alpha} \frac{\omega_{\rm d}^2}{\omega_{\alpha}^2} \bepsilon_{\alpha}\cdot\left(\Delta \hat{\bJ}_{\rm p} - \Delta \hat{\bJ}_{\rm d} \right),   
\end{align}
where $ \hat{\bJ}_{\rm p} = -\imagi \hbar\sum_i \nabla_i$ and $\hat{\bJ}_{\rm d} = \frac{q}{c}N_e  \hat{\bA}$. 
Eq.~\ref{eq:varlimit} can then be solved for $\hat{\bA}$ by the Bogoliubov transformation introduced in App.~\ref{App:bogo} which leads to the 
new frequencies $\tilde{\omega}_{\alpha}$ and polarization vectors $\tilde{\bepsilon}_{\alpha}$. If we further allow to take into account a cavity in the long-wavelength limit by changing the $\lambda_{\alpha}$ and $\bepsilon_{\alpha}$, only keep a few effective modes, subsume the rest of the modes in the physical mass of the electrons $m \rightarrow m_{\rm e}$ and use atomic units ($\hbar=|e|=m_{\rm e}=1/(4 \pi \epsilon_0) =1$), we recover exactly the case of the main text. We note that for the sake of consistency, if we change the local form of the modes by hand, also the longitudinal modes will change as can be seen from Eq.~\eqref{eq:CoulombModes}. We therefore use a generic $w(\br,\br')$ in the main text to accommodate also this eventuality.

This connection to the long-wavelength limit also directly shows that if we go beyond the dipole approximation, in lowest order we can just re-substitute $\lambda_{\alpha} \rightarrow \lambda_{\alpha}(\br)$ and $\bepsilon_{\alpha} \rightarrow \bepsilon_{\alpha}(\br)$. 

\section{Time-dependent photon-exchange approximation}
\label{App:tdMxapproximation}

In the time-dependent case the simple idea that the interacting and the auxiliary system have both the same (zero) paramagnetic currents does no longer hold. Various choices for the basic variables of QEDFT are possible~\cite{tokatly2013, ruggenthaler2014}. The one that is most consistent with our static discussion is to make the physical currents the same in both systems. This is then a more general setting for the long-wavelength situation than the usual density-based QEDFT~\cite{tokatly2013,ruggenthaler2014}. 

We therefore use the orbital equations (here again in atomic units)
\begin{align}\label{eq:tdKS}
    \imagi \partial_t \varphi_i(\br,t) = \left[ \frac{1}{2}\left(-\imagi \nabla + \frac{1}{c}\bA_{s}(\br,t)  \right)^2 + v_s(\br,t)\right]\varphi_i(\br,t)~,
\end{align}
for the Kohn-Sham system,
whereas in the static case we only had the spatially independent $\bA_s = \bA$. In the static case this homogeneous effective field can be discarded without loss of generality. It corresponds to a trivial global gauge transformation.
For the interacting system the basic equation of motion for the matter is~\cite{ruggenthaler2014, jestadt2018real}
\begin{align}\label{eq:tdEOM}
    &\frac{\rmd}{\rmd t} \bj(\br,t) = \FF_{T}([\Psi],\br,t) + \FF_{W}([\Psi],\br,t) \\
    &-\frac{1}{c}\left\langle \big(\hat{\bA}_{\rm H}(t) \cdot \nabla \big) \hat{\mathbf{j}}_\mathrm{p,H}(\br,t) \right\rangle - \rho(\br,t)\nabla v(\br,t) \nonumber \\
    & - \frac{1}{c} \left\langle \hat{\bA}_{\rm H}(t) \big( \nabla \cdot \hat{\bj}_{\rm H}(\br,t) \big) \right\rangle - \left\langle \hat{\mathbf{E}}_{\rm H}(t) \hat{\rho}_{\rm H}(\br, t)\right \rangle, \nonumber
\end{align}
where the physical current in the Schr\"odinger picture is $\hat{\bj}(\br) = \hat{\bj}_{\rm p}(\br ) + \tfrac{N_e}{c}\hat{\bA}$ and $-\tfrac{1}{c}\partial_t \hat{\bA}_{\rm H}(t) = \hat{\mathbf{E}}_{\rm H}(t)$. Here $\langle \cdot \rangle$ indicates to evaluate the expectation value with a fixed initial state of the coupled light-matter system $\Psi$, which will usually be the ground state of the Pauli-Fierz Hamiltonian Eq.~\eqref{eq:PF_Hamiltonian}. The corresponding equation of motion for the Kohn-Sham system is then (see also discussion in App.~\ref{App:mcMxapproximation}, while the Stern-Gerlach part is now omitted in accordance with Eq.~\eqref{eq:tdKS})
\begin{align*}
    &\frac{\rmd}{\rmd t} j^k(\br,t) = F^{k}_{T}([\Phi],\br,t) \\
    & - \frac{1}{c} \sum_{l=1}^3 \left[ A^{l}_{\rm s}(\br,t) \partial_l j^k_{\rm p}([\Phi],\br,t) + \left(\partial_k A^{l}_{\rm s}(\br,t)\right) j^{l}_{\rm p}([\Phi],\br,t)  \right] \nonumber \\
    &-\left[ \partial_k \left(\frac{1}{2  c^2}\bA_{\rm s}(\br,t)^2 + v_s(\br,t)  \right)\right]\rho(\br,t) \nonumber \\
    & + \frac{1}{c} \left(\partial_t A^{k}_{\rm s}(\br,t)\right)\rho(\br,t) - \frac{1}{c}A^{k}_{\rm s}(\br,t)\nabla \cdot \bj(\br,t)~,
\end{align*}
where we have already used that it generates the same density and current density as the interacting reference system. In general, we would also have contributions due to an external vector potential in Coulomb gauge $\bA_{\rm ext}(\br,t)$ in Eq.~\eqref{eq:tdEOM}, which is then used to derive the basic mapping theorems~\cite{tokatly2013, ruggenthaler2014}. But since we are only interested in the case where $\bA_{\rm ext} \equiv 0$ here, we have skipped this possibility from the start for notational simplicity. 

We can then define the Mxc potentials via
\begin{align}\label{eq:FullMHxc}
    \partial_k v_{\rm Mxc} &- \frac{1}{c} \partial_t A^k_{\rm Mxc} = \frac{1}{\rho} \bigg[ F^k_{T}[\Phi] - F^k_{T}[\Psi] - F^{k}_{W}[\Psi] \nonumber
    \\
    & - \frac{1}{c} \sum_{l=1}^3 \left( A^{l}_{\rm Mxc} \partial_l j^k_{\rm p}[\Phi] + \left(\partial_k A^{l}_{\rm Mxc}\right) j^{l} \right) \nonumber
    \\
    & -\frac{1}{c}\sum_{l=1}^{3} A^k_{\rm Mxc} \partial_l j^l +\frac{1}{c}\sum_{l=1}^{3}\left\langle \hat{A}^l_{\rm H} \partial_l \hat{j}_{\rm p,H}^k \right\rangle 
    \nonumber \\
    &+ \frac{1}{c} \sum_{l=1}^{3} \left\langle \hat{A}^k_{\rm H} \partial_l \hat{j}^{l}_{\rm H}\right\rangle + \left\langle \hat{E}^k_{\rm H} \hat{\rho}_{\rm H}\right \rangle \bigg]~,
\end{align}
where we used $v_{\rm Mxc} = v_s - v$ and $\bA_{\rm Mxc} = \bA_{\rm s} - \bA_{\rm ext} $. Now, denoting the right-hand side of Eq.~\eqref{eq:FullMHxc} by $Q^{k}[\Phi,\Psi]$, we can use the Helmholtz decomposition to find in accordance with Eq.~\eqref{eq:staticHxpotential} the (longitudinal) scalar Mxc potential
\begin{align}\label{eq:app:staticMHxpotential}
    \nabla^2 v_{\rm Mxc}(\br,t) = \nabla \cdot \boldsymbol{Q}([\Phi,\Psi],\br,t)
\end{align}
and the (transverse) vector Mxc potential
\begin{align*}
    -\partial_t \bA_{\rm Mxc}(\br,t) = c \boldsymbol{Q}_{\perp}([\Phi,\Psi], \br,t)~.
\end{align*}
In the static case, where the paramagnetic and diamagnetic contributions are individually zero, all the diamagnetic parts cancel and $\bA_{\rm Mxc} \equiv 0$ and we are left with  Eq.~\eqref{eq:app:staticMHxpotential} only.

Again, we need to find an approximation to the photonic part in terms of Kohn-Sham quantities. As a first step we rewrite $\bA_{\rm Mxc}(\br,t) = \bA(t) + \bA_{\rm xc}(\br,t)$ such that we can shift $\bA(t)$ to form the fluctuation operator $\Delta \hat{\bA}$ in Eq.~\eqref{eq:FullMHxc}. Then we follow the strategy of the static case and employ the mode-resolved inhomogeneous Maxwell's equation for $\hat{A}^k = \sum_{\alpha} \hat{A}_{\alpha}\tilde{\epsilon}_\alpha^k$, which becomes
\begin{align*}
    \left(\partial_t^2 + \tilde{\omega}_{\alpha}^2  \right)\hat{A}_{\alpha, \rm H}(t) = -\frac{c \omega_{\alpha, \mathrm{d}}^2}{N}\hat{\bJ}_{\rm p, H}(t)\cdot \tilde{\bepsilon}_{\alpha}~.
\end{align*}
This can be solved formally by
\begin{align*}
    \hat{A}_{\alpha, \rm H}(t) = &-\frac{c \omega_{\alpha, \mathrm{d}}^2}{N} \int_0^{t} \mathrm{d} t' \; \frac{\sin(\tilde{\omega}_{\alpha}(t-t'))}{\tilde{\omega}_{\alpha}}\hat{\bJ}_{\rm p, H}(t')\cdot \tilde{\bepsilon}_{\alpha} \nonumber \\
    &+ \hat{A}_{\alpha} \cos(\tilde{\omega}_{\alpha} t) + \frac{\partial_t \hat{A}_{\alpha, \rm H}(0)}{\tilde{\omega}_{\alpha}} \sin(\tilde{\omega}_{\alpha} t)~.
\end{align*}
We then define the px approximation by using $\hat{A}_{\alpha, \rm H}(t) = A_{\alpha}(t) + \Delta \hat{A}_{\alpha, \rm H}(t)$ and replacing 
\begin{align*}
    &\Delta \hat{A}_{\alpha} \rightarrow -c \omda^2/(N_e \tilde{\omega}^2_{\alpha}) \tilde{\bepsilon}_{\alpha}\cdot \Delta \hatJp~,\\
    &\Delta \partial_t \hat{A}_{\alpha, \rm H}(0) \rightarrow -c \omda^2/(N_e \tilde{\omega}^2_{\alpha}) \tilde{\bepsilon}_{\alpha}\cdot \Delta \partial_t \hat{\bJ}_{\rm p,H}(0)~, \quad\text{and}\\
    &\Delta \hat{\bJ}_{\rm p,H}(t) \rightarrow \Delta \hat{\bJ}_{\rm p, H_s}(t)~,
\end{align*}
where ${\rm H}_s$ indicates that we use now the Kohn-Sham system Heisenberg picture. Further, we use a symmetrized form of the photonic expressions as discussed in Sec.~\ref{sec:functionals}, evaluate all expectation values with the auxiliary Kohn-Sham wave function, and denote the resulting part of the Mxc vector potential as $\bA_{\rm px}(\br,t) = \bA(t) + \bA_{\rm x}(\br,t)$. This way we find several further terms in the non-adiabatic px approximation. In the static case only the cosine term survives and we recover exactly Eq.~\eqref{eq:staticHxpotential}. 